\documentclass[zpreprint,zbstepj]{zeus_paper}

\usepackage[english]{babel}

\newcommand{\Zctdmvddesc}[1]{%
In the kinematic range of the analysis, charged particles were tracked
in the central tracking detector (CTD)~\citeCTD and, for the data
taken after 2001, also in the microvertex detector
(MVD)~\citeMVD. These components operated in a magnetic field of
$1.43\Tesla$ provided by a thin superconducting solenoid. The CTD
consisted of 72~cylindrical drift chamber layers, organised in nine
superlayers covering the polar-angle region
\mbox{$15^\circ<\theta<164^\circ$}. The MVD provided polar angle coverage from $7^\circ$ 
to $150^\circ$. }

\chardef\usc=95
\chardef\til=126
\catcode`\@=11 
\DeclareRobustCommand\xdotspace{\futurelet\@let@token\@xdotspace}
\def\@xdotspace{%
  \ifx\@let@token.\else
  \ifx\@let@token\bgroup.\else
  \ifx\@let@token\egroup.\else
  \ifx\@let@token\/.\else
  \ifx\@let@token\ .\else
  \ifx\@let@token~.\else
  \ifx\@let@token!.\else
  \ifx\@let@token,.\else
  \ifx\@let@token:.\else
  \ifx\@let@token;.\else
  \ifx\@let@token?.\else
  \ifx\@let@token/.\else
  \ifx\@let@token'.\else
  \ifx\@let@token).\else
  \ifx\@let@token-.\else
  \ifx\@let@token\@xobeysp.\else
  \ifx\@let@token\space.\else
  \ifx\@let@token\@sptoken.\else
   .\space
   \fi\fi\fi\fi\fi\fi\fi\fi\fi\fi\fi\fi\fi\fi\fi\fi\fi\fi}
\catcode`\@=12 

\newcommand{\stru}[2]{%
   \relax\ifmmode\hbox{\vrule height#1 depth#2 width0pt}%
   \else\vrule height#1 depth#2 width0pt\fi}

\newcommand{\Ronum}[1]{\uppercase\expandafter{\romannumeral#1}}
\newcommand{\ronum}[1]{\expandafter{\romannumeral#1}}
\DeclareRobustCommand{\LaTeXZ}{%
  \LaTeX\kern-.05em4\kern-.1em
  {\raisebox{-0.2ex}{$\scriptstyle\text{ZEUS}$}}\xspace}

\DeclareMathAlphabet{\mathbf}{OT1}{cmr}{bx}{sl}
\newcommand{\eVdist}{\kern-0.06667em}

\newcommand{\mev}{{\,\text{Me}\eVdist\text{V\/}}}

\newcommand{\Tesla}{\,\text{T}}

\newcommand{\slashfrac}[2]{%
  \raisebox{0.5ex}{\ensuremath #1}\kern-0.12em/\kern-0.08em
  \raisebox{-.8ex}{\ensuremath #2}}

\newcommand{\sqr}[3]{%
    {\vcenter{\hrule height.#3ex\hbox{\vrule width.#2ex height#1ex
     \kern#1ex\vrule width.#3ex}\hrule height.#2ex}}}

\catcode`\@=11 
\newcommand{\parenbar}{\mathpalette\p@renb@r}
\def\p@renb@r#1#2{\vbox{%
  \ifx#1\scriptscriptstyle \dimen@.7em\dimen@ii.2em\else
  \ifx#1\scriptstyle \dimen@.8em\dimen@ii.25em\else
  \dimen@1em\dimen@ii.4em\fi\fi \offinterlineskip
  \ialign{\hfill##\hfill\cr
    \vbox{\hrule width\dimen@ii}\cr
    \noalign{\vskip-.3ex}%
    \hbox to\dimen@{$\mathchar300\hfil\mathchar301$}\cr
    \noalign{\vskip-.3ex}%
    $#1#2$\cr}}}
\catcode`\@=12 

\newcommand{\IP}{{\rm I$\kern-0.01667em$P}\xspace}

\mathchardef\qsm=63
\mathchardef\pls=43
\mathchardef\mns=512
\mathchardef\plm=518
\mathchardef\eql=61
\mathchardef\smallleft=300
\mathchardef\smallright=301
\mathchardef\les=316
\mathchardef\gre=318
\mathchardef\leq=532
\mathchardef\grq=533

\catcode`\@=11 
\newcounter{pict@width}
\newcounter{pict@height}
\newlength{\pict@scale}
\newcommand{\psfigadd}[4]{%
\setcounter{pict@width}{1*\ratio{#2+\pict@scale/2}{\pict@scale}}
\setcounter{pict@height}{1*\ratio{#3+\pict@scale/2}{\pict@scale}}
\setlength{\unitlength}{\pict@scale}
\hbox to #2{\hspace{-\fill}\begin{picture}(\thepict@width,\thepict@height)
\put(0,0){\psfig{figure=#1,width=#2,height=#3,clip=}}
\SetScale{0.283466457}
\SetWidth{1.763889}
{#4}
\end{picture}}
}
\newcounter{pict@widthfst}
\newcounter{pict@widthscd}
\newcounter{pict@widthtot}
\newcommand{\psfigaddtwo}[7]{%
\setcounter{pict@widthfst}{1*\ratio{#2+\pict@scale/2}{\pict@scale}}
\setcounter{pict@widthscd}{1*\ratio{#2+#4+\pict@scale/2}{\pict@scale}}
\setcounter{pict@widthtot}{1*\ratio{#2+#4+#6+\pict@scale/2}{\pict@scale}}
\setcounter{pict@height}{1*\ratio{#3+\pict@scale/2}{\pict@scale}}
\setlength{\unitlength}{\pict@scale}
\hbox{\hspace{-\fill}\begin{picture}(\thepict@widthtot,\thepict@height)
\put(0,0){\psfig{figure=#1,width=#2,height=#3,clip=}}
\put(\thepict@widthscd,0){\psfig{figure=#5,width=#6,height=#3,clip=}}
\SetScale{0.283466457}
\SetWidth{1.763889}
{#7}
\end{picture}}
}
\newcommand{\psfigror}[4]{%
\setcounter{pict@width}{1*\ratio{#2+\pict@scale/2}{\pict@scale}}
\setcounter{pict@height}{1*\ratio{#3+\pict@scale/2}{\pict@scale}}
\setlength{\unitlength}{\pict@scale}
\hbox{\begin{picture}(\thepict@width,\thepict@height)
\put(0,\thepict@height){\psfig{figure=#1,width=#3,height=#2,clip=,angle=270}}
\SetScale{0.283466457}
\SetWidth{1.763889}
{#4}
\end{picture}}
}
\newcommand{\psfigrol}[4]{%
\setcounter{pict@width}{1*\ratio{#2+\pict@scale/2}{\pict@scale}}
\setcounter{pict@height}{1*\ratio{#3+\pict@scale/2}{\pict@scale}}
\setlength{\unitlength}{\pict@scale}
\hbox{\begin{picture}(\thepict@width,\thepict@height)
\put(0,0){\psfig{figure=#1,width=#3,height=#2,clip=,angle=90}}
\SetScale{0.283466457}
\SetWidth{1.763889}
{#4}
\end{picture}}
}
\catcode`\@=12 
\newlength\listtextwidth

\catcode`\@=11 
\newlength{\@tabfninsert}
\newlength{\@tabfnwidth}
\newcommand{\tabfootnote}[2]{%
  \setlength{\@tabfninsert}{0.8em}
  \setlength{\@tabfnwidth}{\textwidth}
  \addtolength{\@tabfnwidth}{-\@tabfninsert}
  \addtolength{\@tabfnwidth}{-0.4em}
  \noindent\makebox[\@tabfninsert][r]{\footnotesize$^{#1}$\hfil}\hfill%
  \parbox[t]{\@tabfnwidth}{\footnotesize #2\hfill}}
\catcode`\@=12 

\newcommand{\be}{\begin{eqnarray}}
\newcommand{\ee}{\end{eqnarray}}
\newcommand{\mpp}{M_{\pi\pi}}

\newcommand{\rh}{\rho}
\newcommand{\rhp}{\rho^\prime}
\newcommand{\rhpp}{\rho^{\prime\prime}}
\newcommand{\ffpi}{|F_{\pi}(M_{\pi\pi})|^2}
\newcommand{\fpi}{F_{\pi}(M_{\pi\pi})}

\def\beq{\begin{equation}}
\def\eeq{\end{equation}}

\newcommand\units{\,\,\mathrm}
\newcommand{\gevu}{\units{GeV}}

\begin{document}

\prepnum{{DESY--11--220}}

\title{Exclusive electroproduction of two pions at HERA}                                                       
                    
\author{ZEUS Collaboration}
\draftversion{post-reading}
\date{November 2011}

\abstract{
The exclusive electroproduction of two pions in the mass range
\mbox{$0.4<M_{\pi\pi}<2.5 \gevu$}  has been studied with the ZEUS
detector at HERA using an integrated luminosity of \mbox{82
pb$^{-1}$}.  The analysis was carried out in the kinematic range of
$2<Q^2<80 \gevu^2$, $32<W<180 \gevu$ and $|t|<0.6
\gevu^2$, where $Q^2$ is the photon virtuality, $W$ is the
photon-proton centre-of-mass energy and $t$ is the squared
four-momentum transfer at the proton vertex.  The two-pion
invariant-mass distribution is interpreted in terms of the pion
electromagnetic form factor, $|F(M_{\pi\pi})|$, assuming that the
studied mass range includes the contributions of the $\rh,\
\rhp$ and $\rhpp$ vector-meson states.  The masses and widths of
the resonances were obtained and the $Q^2$ dependence of the
cross-section ratios $\sigma(\rhp\to\pi\pi)/\sigma(\rh)$ and
$\sigma(\rhpp\to\pi\pi)/\sigma(\rh)$ was extracted. The pion form
factor obtained in the present analysis is compared to that obtained
in $e^+e^-\to\pi^+\pi^-$.

  }

\makezeustitle

\newpage

\def\3{\ss}
\pagenumbering{Roman}

\begin{center}
{                      \Large  The ZEUS Collaboration              }
\end{center}

{\small


        {\raggedright
H.~Abramowicz$^{45, ah}$, 
I.~Abt$^{35}$, 
L.~Adamczyk$^{13}$, 
M.~Adamus$^{54}$, 
R.~Aggarwal$^{7, c}$, 
S.~Antonelli$^{4}$, 
P.~Antonioli$^{3}$, 
A.~Antonov$^{33}$, 
M.~Arneodo$^{50}$, 
D.~Ashery$^{45}$
% ...
        }

\newpage


\makebox[3em]{$^{1}$}
\begin{minipage}[t]{14cm}
{\it Argonne National Laboratory, Argonne, Illinois 60439-4815, USA}~$^{A}$

\end{minipage}\\
\makebox[3em]{$^{2}$}
\begin{minipage}[t]{14cm}
{\it Andrews University, Berrien Springs, Michigan 49104-0380, USA}

\end{minipage}\\
\makebox[3em]{$^{3}$}
\begin{minipage}[t]{14cm}
{\it INFN Bologna, Bologna, Italy}~$^{B}$

\end{minipage}\\
\makebox[3em]{$^{4}$}
\begin{minipage}[t]{14cm}
{\it University and INFN Bologna, Bologna, Italy}~$^{B}$

\end{minipage}\\
\makebox[3em]{$^{5}$}
\begin{minipage}[t]{14cm}
{\it Physikalisches Institut der Universit\"at Bonn,
Bonn, Germany}~$^{C}$

\end{minipage}\\
\makebox[3em]{$^{6}$}
\begin{minipage}[t]{14cm}
{\it H.H.~Wills Physics Laboratory, University of Bristol,
Bristol, United Kingdom}~$^{D}$

\end{minipage}\\
\makebox[3em]{$^{7}$}
\begin{minipage}[t]{14cm}
{\it Panjab University, Department of Physics, Chandigarh, India}

\end{minipage}\\
\makebox[3em]{$^{8}$}
\begin{minipage}[t]{14cm}
{\it Calabria University,
Physics Department and INFN, Cosenza, Italy}~$^{B}$

\end{minipage}\\
\makebox[3em]{$^{9}$}
\begin{minipage}[t]{14cm}
{\it Institute for Universe and Elementary Particles, Chonnam National University,\\
Kwangju, South Korea}

\end{minipage}\\
\makebox[3em]{$^{10}$}
\begin{minipage}[t]{14cm}
{\it Jabatan Fizik, Universiti Malaya, 50603 Kuala Lumpur, Malaysia}~$^{E}$

\end{minipage}\\
\makebox[3em]{$^{11}$}
\begin{minipage}[t]{14cm}
{\it Nevis Laboratories, Columbia University, Irvington on Hudson,
New York 10027, USA}~$^{F}$

\end{minipage}\\
\makebox[3em]{$^{12}$}
\begin{minipage}[t]{14cm}
{\it The Henryk Niewodniczanski Institute of Nuclear Physics, Polish Academy of \\
Sciences, Krakow, Poland}~$^{G}$

\end{minipage}\\
\makebox[3em]{$^{13}$}
\begin{minipage}[t]{14cm}
{\it AGH-University of Science and Technology, Faculty of Physics and Applied Computer
Science, Krakow, Poland}~$^{H}$

\end{minipage}\\
\makebox[3em]{$^{14}$}
\begin{minipage}[t]{14cm}
{\it Department of Physics, Jagellonian University, Cracow, Poland}

\end{minipage}\\
\makebox[3em]{$^{15}$}
\begin{minipage}[t]{14cm}
{\it Deutsches Elektronen-Synchrotron DESY, Hamburg, Germany}

\end{minipage}\\
\makebox[3em]{$^{16}$}
\begin{minipage}[t]{14cm}
{\it Deutsches Elektronen-Synchrotron DESY, Zeuthen, Germany}

\end{minipage}\\
\makebox[3em]{$^{17}$}
\begin{minipage}[t]{14cm}
{\it INFN Florence, Florence, Italy}~$^{B}$

\end{minipage}\\
\makebox[3em]{$^{18}$}
\begin{minipage}[t]{14cm}
{\it University and INFN Florence, Florence, Italy}~$^{B}$

\end{minipage}\\
\makebox[3em]{$^{19}$}
\begin{minipage}[t]{14cm}
{\it Fakult\"at f\"ur Physik der Universit\"at Freiburg i.Br.,
Freiburg i.Br., Germany}

\end{minipage}\\
\makebox[3em]{$^{20}$}
\begin{minipage}[t]{14cm}
{\it School of Physics and Astronomy, University of Glasgow,
Glasgow, United Kingdom}~$^{D}$

\end{minipage}\\
\makebox[3em]{$^{21}$}
\begin{minipage}[t]{14cm}
{\it Department of Engineering in Management and Finance, Univ. of
the Aegean, Chios, Greece}

\end{minipage}\\
\makebox[3em]{$^{22}$}
\begin{minipage}[t]{14cm}
{\it Hamburg University, Institute of Experimental Physics, Hamburg,
Germany}~$^{I}$

\end{minipage}\\
\makebox[3em]{$^{23}$}
\begin{minipage}[t]{14cm}
{\it Imperial College London, High Energy Nuclear Physics Group,
London, United Kingdom}~$^{D}$

\end{minipage}\\
\makebox[3em]{$^{24}$}
\begin{minipage}[t]{14cm}
{\it Institute of Particle and Nuclear Studies, KEK,
Tsukuba, Japan}~$^{J}$

\end{minipage}\\
\makebox[3em]{$^{25}$}
\begin{minipage}[t]{14cm}
{\it Institute of Physics and Technology of Ministry of Education and
Science of Kazakhstan, Almaty, Kazakhstan}

\end{minipage}\\
\makebox[3em]{$^{26}$}
\begin{minipage}[t]{14cm}
{\it Institute for Nuclear Research, National Academy of Sciences, Kyiv, Ukraine}

\end{minipage}\\
\makebox[3em]{$^{27}$}
\begin{minipage}[t]{14cm}
{\it Department of Nuclear Physics, National Taras Shevchenko University of Kyiv, Kyiv, Ukraine}

\end{minipage}\\
\makebox[3em]{$^{28}$}
\begin{minipage}[t]{14cm}
{\it Kyungpook National University, Center for High Energy Physics, Daegu,
South Korea}~$^{K}$

\end{minipage}\\
\makebox[3em]{$^{29}$}
\begin{minipage}[t]{14cm}
{\it Institut de Physique Nucl\'{e}aire, Universit\'{e} Catholique de Louvain, Louvain-la-Neuve,\\
Belgium}~$^{L}$

\end{minipage}\\
\makebox[3em]{$^{30}$}
\begin{minipage}[t]{14cm}
{\it Departamento de F\'{\i}sica Te\'orica, Universidad Aut\'onoma
de Madrid, Madrid, Spain}~$^{M}$

\end{minipage}\\
\makebox[3em]{$^{31}$}
\begin{minipage}[t]{14cm}
{\it Department of Physics, McGill University,
Montr\'eal, Qu\'ebec, Canada H3A 2T8}~$^{N}$

\end{minipage}\\
\makebox[3em]{$^{32}$}
\begin{minipage}[t]{14cm}
{\it Meiji Gakuin University, Faculty of General Education,
Yokohama, Japan}~$^{J}$

\end{minipage}\\
\makebox[3em]{$^{33}$}
\begin{minipage}[t]{14cm}
{\it Moscow Engineering Physics Institute, Moscow, Russia}~$^{O}$

\end{minipage}\\
\makebox[3em]{$^{34}$}
\begin{minipage}[t]{14cm}
{\it Moscow State University, Institute of Nuclear Physics,
Moscow, Russia}~$^{P}$

\end{minipage}\\
\makebox[3em]{$^{35}$}
\begin{minipage}[t]{14cm}
{\it Max-Planck-Institut f\"ur Physik, M\"unchen, Germany}

\end{minipage}\\
\makebox[3em]{$^{36}$}
\begin{minipage}[t]{14cm}
{\it NIKHEF and University of Amsterdam, Amsterdam, Netherlands}~$^{Q}$

\end{minipage}\\
\makebox[3em]{$^{37}$}
\begin{minipage}[t]{14cm}
{\it Physics Department, Ohio State University,
Columbus, Ohio 43210, USA}~$^{A}$

\end{minipage}\\
\makebox[3em]{$^{38}$}
\begin{minipage}[t]{14cm}
{\it Department of Physics, University of Oxford,
Oxford, United Kingdom}~$^{D}$

\end{minipage}\\
\makebox[3em]{$^{39}$}
\begin{minipage}[t]{14cm}
{\it INFN Padova, Padova, Italy}~$^{B}$

\end{minipage}\\
\makebox[3em]{$^{40}$}
\begin{minipage}[t]{14cm}
{\it Dipartimento di Fisica dell' Universit\`a and INFN,
Padova, Italy}~$^{B}$

\end{minipage}\\
\makebox[3em]{$^{41}$}
\begin{minipage}[t]{14cm}
{\it Department of Physics, Pennsylvania State University, University Park,\\
Pennsylvania 16802, USA}~$^{F}$

\end{minipage}\\
\makebox[3em]{$^{42}$}
\begin{minipage}[t]{14cm}
{\it Polytechnic University, Sagamihara, Japan}~$^{J}$

\end{minipage}\\
\makebox[3em]{$^{43}$}
\begin{minipage}[t]{14cm}
{\it Dipartimento di Fisica, Universit\`a 'La Sapienza' and INFN,
Rome, Italy}~$^{B}$

\end{minipage}\\
\makebox[3em]{$^{44}$}
\begin{minipage}[t]{14cm}
{\it Rutherford Appleton Laboratory, Chilton, Didcot, Oxon,
United Kingdom}~$^{D}$

\end{minipage}\\
\makebox[3em]{$^{45}$}
\begin{minipage}[t]{14cm}
{\it Raymond and Beverly Sackler Faculty of Exact Sciences, School of Physics, \\
Tel Aviv University, Tel Aviv, Israel}~$^{R}$

\end{minipage}\\
\makebox[3em]{$^{46}$}
\begin{minipage}[t]{14cm}
{\it Department of Physics, Tokyo Institute of Technology,
Tokyo, Japan}~$^{J}$

\end{minipage}\\
\makebox[3em]{$^{47}$}
\begin{minipage}[t]{14cm}
{\it Department of Physics, University of Tokyo,
Tokyo, Japan}~$^{J}$

\end{minipage}\\
\makebox[3em]{$^{48}$}
\begin{minipage}[t]{14cm}
{\it Tokyo Metropolitan University, Department of Physics,
Tokyo, Japan}~$^{J}$

\end{minipage}\\
\makebox[3em]{$^{49}$}
\begin{minipage}[t]{14cm}
{\it Universit\`a di Torino and INFN, Torino, Italy}~$^{B}$

\end{minipage}\\
\makebox[3em]{$^{50}$}
\begin{minipage}[t]{14cm}
{\it Universit\`a del Piemonte Orientale, Novara, and INFN, Torino,
Italy}~$^{B}$

\end{minipage}\\
\makebox[3em]{$^{51}$}
\begin{minipage}[t]{14cm}
{\it Department of Physics, University of Toronto, Toronto, Ontario,
Canada M5S 1A7}~$^{N}$

\end{minipage}\\
\makebox[3em]{$^{52}$}
\begin{minipage}[t]{14cm}
{\it Physics and Astronomy Department, University College London,
London, United Kingdom}~$^{D}$

\end{minipage}\\
\makebox[3em]{$^{53}$}
\begin{minipage}[t]{14cm}
{\it Faculty of Physics, University of Warsaw, Warsaw, Poland}

\end{minipage}\\
\makebox[3em]{$^{54}$}
\begin{minipage}[t]{14cm}
{\it National Centre for Nuclear Research, Warsaw, Poland}

\end{minipage}\\
\makebox[3em]{$^{55}$}
\begin{minipage}[t]{14cm}
{\it Department of Particle Physics and Astrophysics, Weizmann
Institute, Rehovot, Israel}

\end{minipage}\\
\makebox[3em]{$^{56}$}
\begin{minipage}[t]{14cm}
{\it Department of Physics, University of Wisconsin, Madison,
Wisconsin 53706, USA}~$^{A}$

\end{minipage}\\
\makebox[3em]{$^{57}$}
\begin{minipage}[t]{14cm}
{\it Department of Physics, York University, Ontario, Canada M3J
1P3}~$^{N}$

\end{minipage}\\
\vspace{30em} \pagebreak[4]


\makebox[3ex]{$^{ A}$}
\begin{minipage}[t]{14cm}
 supported by the US Department of Energy\
\end{minipage}\\
\makebox[3ex]{$^{ B}$}
\begin{minipage}[t]{14cm}
 supported by the Italian National Institute for Nuclear Physics (INFN) \
\end{minipage}\\
\makebox[3ex]{$^{ C}$}
\begin{minipage}[t]{14cm}
 supported by the German Federal Ministry for Education and Research (BMBF), under
 contract No. 05 H09PDF\
\end{minipage}\\
\makebox[3ex]{$^{ D}$}
\begin{minipage}[t]{14cm}
 supported by the Science and Technology Facilities Council, UK\
\end{minipage}\\
\makebox[3ex]{$^{ E}$}
\begin{minipage}[t]{14cm}
 supported by an FRGS grant from the Malaysian government\
\end{minipage}\\
\makebox[3ex]{$^{ F}$}
\begin{minipage}[t]{14cm}
 supported by the US National Science Foundation. Any opinion,
 findings and conclusions or recommendations expressed in this material
 are those of the authors and do not necessarily reflect the views of the
 National Science Foundation.\
\end{minipage}\\
\makebox[3ex]{$^{ G}$}
\begin{minipage}[t]{14cm}
 supported by the Polish Ministry of Science and Higher Education as a scientific project No.
 DPN/N188/DESY/2009\
\end{minipage}\\
\makebox[3ex]{$^{ H}$}
\begin{minipage}[t]{14cm}
 supported by the Polish Ministry of Science and Higher Education and its grants
 for Scientific Research\
\end{minipage}\\
\makebox[3ex]{$^{ I}$}
\begin{minipage}[t]{14cm}
 supported by the German Federal Ministry for Education and Research (BMBF), under
 contract No. 05h09GUF, and the SFB 676 of the Deutsche Forschungsgemeinschaft (DFG) \
\end{minipage}\\
\makebox[3ex]{$^{ J}$}
\begin{minipage}[t]{14cm}
 supported by the Japanese Ministry of Education, Culture, Sports, Science and Technology
 (MEXT) and its grants for Scientific Research\
\end{minipage}\\
\makebox[3ex]{$^{ K}$}
\begin{minipage}[t]{14cm}
 supported by the Korean Ministry of Education and Korea Science and Engineering
 Foundation\
\end{minipage}\\
\makebox[3ex]{$^{ L}$}
\begin{minipage}[t]{14cm}
 supported by FNRS and its associated funds (IISN and FRIA) and by an Inter-University
 Attraction Poles Programme subsidised by the Belgian Federal Science Policy Office\
\end{minipage}\\
\makebox[3ex]{$^{ M}$}
\begin{minipage}[t]{14cm}
 supported by the Spanish Ministry of Education and Science through funds provided by
 CICYT\
\end{minipage}\\
\makebox[3ex]{$^{ N}$}
\begin{minipage}[t]{14cm}
 supported by the Natural Sciences and Engineering Research Council of Canada (NSERC) \
\end{minipage}\\
\makebox[3ex]{$^{ O}$}
\begin{minipage}[t]{14cm}
 partially supported by the German Federal Ministry for Education and Research (BMBF)\
\end{minipage}\\
\makebox[3ex]{$^{ P}$}
\begin{minipage}[t]{14cm}
 supported by RF Presidential grant N 4142.2010.2 for Leading Scientific Schools, by the
 Russian Ministry of Education and Science through its grant for Scientific Research on
 High Energy Physics and under contract No.02.740.11.0244 \
\end{minipage}\\
\makebox[3ex]{$^{ Q}$}
\begin{minipage}[t]{14cm}
 supported by the Netherlands Foundation for Research on Matter (FOM)\
\end{minipage}\\
\makebox[3ex]{$^{ R}$}
\begin{minipage}[t]{14cm}
 supported by the Israel Science Foundation\
\end{minipage}\\
\vspace{30em} \pagebreak[4]


\makebox[3ex]{$^{ a}$}
\begin{minipage}[t]{14cm}
now at University of Salerno, Italy\
\end{minipage}\\
\makebox[3ex]{$^{ b}$}
\begin{minipage}[t]{14cm}
now at Queen Mary University of London, United Kingdom\
\end{minipage}\\
\makebox[3ex]{$^{ c}$}
\begin{minipage}[t]{14cm}
also funded by Max Planck Institute for Physics, Munich, Germany\
\end{minipage}\\
\makebox[3ex]{$^{ d}$}
\begin{minipage}[t]{14cm}
also Senior Alexander von Humboldt Research Fellow at Hamburg University,
 Institute of Experimental Physics, Hamburg, Germany\
\end{minipage}\\
\makebox[3ex]{$^{ e}$}
\begin{minipage}[t]{14cm}
also at Cracow University of Technology, Faculty of Physics,
 Mathemathics and Applied Computer Science, Poland\
\end{minipage}\\
\makebox[3ex]{$^{ f}$}
\begin{minipage}[t]{14cm}
supported by the research grant No. 1 P03B 04529 (2005-2008)\
\end{minipage}\\
\makebox[3ex]{$^{ g}$}
\begin{minipage}[t]{14cm}
now at Rockefeller University, New York, NY
 10065, USA\
\end{minipage}\\
\makebox[3ex]{$^{ h}$}
\begin{minipage}[t]{14cm}
now at DESY group FS-CFEL-1\
\end{minipage}\\
\makebox[3ex]{$^{ i}$}
\begin{minipage}[t]{14cm}
now at Institute of High Energy Physics, Beijing, China\
\end{minipage}\\
\makebox[3ex]{$^{ j}$}
\begin{minipage}[t]{14cm}
now at DESY group FEB, Hamburg, Germany\
\end{minipage}\\
\makebox[3ex]{$^{ k}$}
\begin{minipage}[t]{14cm}
also at Moscow State University, Russia\
\end{minipage}\\
\makebox[3ex]{$^{ l}$}
\begin{minipage}[t]{14cm}
now at University of Liverpool, United Kingdom\
\end{minipage}\\
\makebox[3ex]{$^{ m}$}
\begin{minipage}[t]{14cm}
now at CERN, Geneva, Switzerland\
\end{minipage}\\
\makebox[3ex]{$^{ n}$}
\begin{minipage}[t]{14cm}
also affiliated with Universtiy College London, UK\
\end{minipage}\\
\makebox[3ex]{$^{ o}$}
\begin{minipage}[t]{14cm}
now at Goldman Sachs, London, UK\
\end{minipage}\\
\makebox[3ex]{$^{ p}$}
\begin{minipage}[t]{14cm}
also at Institute of Theoretical and Experimental Physics, Moscow, Russia\
\end{minipage}\\
\makebox[3ex]{$^{ q}$}
\begin{minipage}[t]{14cm}
also at FPACS, AGH-UST, Cracow, Poland\
\end{minipage}\\
\makebox[3ex]{$^{ r}$}
\begin{minipage}[t]{14cm}
partially supported by Warsaw University, Poland\
\end{minipage}\\
\makebox[3ex]{$^{ s}$}
\begin{minipage}[t]{14cm}
now at Istituto Nucleare di Fisica Nazionale (INFN), Pisa, Italy\
\end{minipage}\\
\makebox[3ex]{$^{ t}$}
\begin{minipage}[t]{14cm}
now at Haase Energie Technik AG, Neum\"unster, Germany\
\end{minipage}\\
\makebox[3ex]{$^{ u}$}
\begin{minipage}[t]{14cm}
now at Department of Physics, University of Bonn, Germany\
\end{minipage}\\
\makebox[3ex]{$^{ v}$}
\begin{minipage}[t]{14cm}
now at Biodiversit\"at und Klimaforschungszentrum (BiK-F), Frankfurt, Germany\
\end{minipage}\\
\makebox[3ex]{$^{ w}$}
\begin{minipage}[t]{14cm}
also affiliated with DESY, Germany\
\end{minipage}\\
\makebox[3ex]{$^{ x}$}
\begin{minipage}[t]{14cm}
also at University of Tokyo, Japan\
\end{minipage}\\
\makebox[3ex]{$^{ y}$}
\begin{minipage}[t]{14cm}
now at Kobe University, Japan\
\end{minipage}\\
\makebox[3ex]{$^{ z}$}
\begin{minipage}[t]{14cm}
supported by DESY, Germany\
\end{minipage}\\
\makebox[3ex]{$^{\dagger}$}
\begin{minipage}[t]{14cm}
 deceased \
\end{minipage}\\
\makebox[3ex]{$^{aa}$}
\begin{minipage}[t]{14cm}
member of National Technical University of Ukraine, Kyiv Polytechnic Institute,
 Kyiv, Ukraine\
\end{minipage}\\
\makebox[3ex]{$^{ab}$}
\begin{minipage}[t]{14cm}
member of National University of Kyiv - Mohyla Academy, Kyiv, Ukraine\
\end{minipage}\\
\makebox[3ex]{$^{ac}$}
\begin{minipage}[t]{14cm}
partly supported by the Russian Foundation for Basic Research, grant 11-02-91345-DFG\_a\
\end{minipage}\\
\makebox[3ex]{$^{ad}$}
\begin{minipage}[t]{14cm}
Alexander von Humboldt Professor; also at DESY and University of
 Oxford\
\end{minipage}\\
\makebox[3ex]{$^{ae}$}
\begin{minipage}[t]{14cm}
STFC Advanced Fellow\
\end{minipage}\\
\makebox[3ex]{$^{af}$}
\begin{minipage}[t]{14cm}
nee Korcsak-Gorzo\
\end{minipage}\\
\makebox[3ex]{$^{ag}$}
\begin{minipage}[t]{14cm}
This material was based on work supported by the
 National Science Foundation, while working at the Foundation.\
\end{minipage}\\
\makebox[3ex]{$^{ah}$}
\begin{minipage}[t]{14cm}
also at Max Planck Institute for Physics, Munich, Germany, External Scientific Member\
\end{minipage}\\
\makebox[3ex]{$^{ai}$}
\begin{minipage}[t]{14cm}
now at Tokyo Metropolitan University, Japan\
\end{minipage}\\
\makebox[3ex]{$^{aj}$}
\begin{minipage}[t]{14cm}
now at Nihon Institute of Medical Science, Japan\
\end{minipage}\\
\makebox[3ex]{$^{ak}$}
\begin{minipage}[t]{14cm}
now at Osaka University, Osaka, Japan\
\end{minipage}\\
\makebox[3ex]{$^{al}$}
\begin{minipage}[t]{14cm}
also at \L\'{o}d\'{z} University, Poland\
\end{minipage}\\
\makebox[3ex]{$^{am}$}
\begin{minipage}[t]{14cm}
member of \L\'{o}d\'{z} University, Poland\
\end{minipage}\\
\makebox[3ex]{$^{an}$}
\begin{minipage}[t]{14cm}
now at Department of Physics, Stockholm University, Stockholm, Sweden\
\end{minipage}\\
\makebox[3ex]{$^{ao}$}
\begin{minipage}[t]{14cm}
also at Cardinal Stefan Wyszy\'nski University, Warsaw, Poland\
\end{minipage}\\

}

\newpage
\pagenumbering{arabic} 
\pagestyle{plain}

\section{Introduction}

Exclusive electroproduction of vector mesons takes place through a
virtual photon $\gamma^*$ by means of the process $\gamma^* p \to V
p$. At large values of the centre-of-mass energy, $W$, this is usually
viewed as a three-step process; the virtual photon $\gamma^*$
fluctuates into a $q\bar{q}$ pair which interacts with the proton
through a two-gluon ladder and hadronizes into a vector meson,
$V$. The production of ground-state vector mesons, $V = \rho,
\omega, \phi, J/\psi, \Upsilon$, which are $1S$ triplet $q\bar{q}$
states, has been extensively studied at HERA, particularly in several
recent
publications~\cite{h1rhophi,zrho,zomega,zphi,h1psi,zpsi,zupsilon}). As
the virtuality, $Q^2$, of the photon increases, the process becomes hard
and can be calculated in perturbative Quantum Chromodynamics
(pQCD). Furthermore, by varying $Q^2$, and thus the size of the
$q\bar{q}$ pair, sensitivity to the vector-meson wave-function can be
obtained by scanning it at different $q\bar{q}$ distance
scales. Expectations in the QCD framework vary from calculations based
only on the mass properties and typical size of the $q\bar{q }$ inside
the vector-meson~\cite{fks96,mrt}, to those based on the details of
the vector meson wave-function dependence on the size of the
$q\bar{q}$ pair~\cite{nnz94,nnpzz98,kdp99,in02,ci05}. The approaches
differ in their predictions for the $Q^2$ dependence of the cross
sections for excited vector-meson states and their ratio to their
ground state.

The only radially excited $2S$ triplet $q\bar{q}$ state studied at
HERA so far has been the $\psi(2S)$ state~\cite{h1psi2s}. In this
study, only the photoproduction reaction was investigated and the low
cross-section ratio of $\psi(2S)$ to the ground-state $J/\psi$
supported the existence of a suppression effect, expected if a node in
the $\psi(2S)$ wave-function is present.

Other excited vector-meson states, in particular those consisting of
light quarks, can be used to study the effect caused by changing the
scanning size. Exclusive $\pi^+\pi^-$ production has been measured
previously in the annihilation process $e^+e^- \to
\pi^+\pi^-$~\cite{barkov}, as well as in photoproduction~\cite{aston}. 
The $\pi^+\pi^-$ mass distribution shows a complex structure in the
mass range 1--2$\gevu$. Evidence for two excited vector-meson states
has been established~\cite{dm87,cd94}; the $\rhp$(1450) is assumed to
be predominantly a radially excited $2S$ state and the $\rhpp$(1700)
is an orbitally excited $2D$ state, with some mixture of the $S$ and
$D$ waves~\cite{in99}. In addition there is also the $\rho_3(1690)$
spin-3 meson~\cite{omega} which has a $\pi\pi$ decay mode.  The two-pion
decay mode of these resonances is related~\cite{p99,cp00} to the pion
electromagnetic form factor, $\fpi$.

In this paper, a study of exclusive electroproduction of two pions,
\begin{equation}
\gamma^* p \to \pi^+ \pi^- p  ,
\label{pipi}
\end{equation}
is presented in the two-pion mass range
\mbox{$0.4<M_{\pi\pi}<2.5\gevu$}, in the kinematic range 
$2<Q^2<80\gevu^2$, $32<W<180\gevu$ and $|t|<0.6\gevu^2$, where $t$
is the squared four-momentum transfer at the proton vertex. The
$M_{\pi\pi}$ system consists of a resonance part and a non-resonant
background. The resonances are described by the pion form factor.  The
contributions of the three vector-mesons $\rho$, $\rhp$ and $\rhpp$
are extracted and their relative rates as a function of $Q^2$ are
discussed in terms of QCD expectations.

\section{The pion form factor}

The two-pion invariant-mass distribution of Eq.~\ref{pipi}, after
subtraction of the non-resonant background\footnote{This is assumed
not to interfere with the resonance signal.}, can be related to the
pion electromagnetic form factor, $\fpi$, through the following
relation~\cite{p99,cp00}:
\begin{equation}
\frac{dN(\mpp)}{d\mpp}\propto |F_{\pi}(\mpp)|^2\, .
\label{mshape}
\end{equation}

There are several parameterizations of the pion form factor usually
used for fitting the $\pi^+\pi^-$ mass distribution; the
Kuhn-Santamaria (KS) ~\cite{kuhn}, the Gounaris-Sakurai
(GS)~\cite{sakurai} and the Hidden Local Symmetry (HLS)
~\cite{hls1,hls2} parameterizations. In this paper, results based on
the KS parameterization are presented.

In the mass range $M_{\pi\pi}<2.5\gevu$, the KS parameterization of
the pion form factor includes contributions from the $\rh$(770),
$\rhp$(1450) and $\rhpp$(1700) resonances\footnote{ This analysis
cannot distinguish between $\rho_3$(1690) and
$\rhpp$(1700). Theoretical calculations estimate the contribution of
$\rho_3$(1690) to be either one order of magnitude~\cite{mrt} or 2--5
times~\cite{ci05-2} smaller than that of the $\rhpp$(1700).},

 \begin{equation}
\fpi = \frac{BW_{\rh}(\mpp) + \beta BW_{\rhp}(\mpp) + \gamma BW_{\rhpp}(\mpp)}
{1+\beta+\gamma} .
\label{fpion}
\end{equation}
Here $\beta$ and $\gamma$ are relative amplitudes and $BW_V$ is the
Breit-Wigner distribution which has the form
\begin{equation}
BW_V(\mpp)=\frac{M_V^2}{M_V^2 - \mpp^2 - i M_V \Gamma_V(M_{\pi\pi})} ,
\label{bw}
\end{equation}
where $M_V$ and $\Gamma_V(M_{\pi\pi})$ are the vector-meson mass and
momentum-dependent width, respectively. The latter has the form
\begin{equation}
 \Gamma_V(M_{\pi\pi})=\Gamma_V \biggl [\frac{p_\pi(M_{\pi\pi})}
{p_\pi(M_{V})}\biggr]^3 
\biggl[\frac{M_{V}^2}{M_{\pi\pi}^2}\biggr], 
\label{width}
\end{equation}
where $\Gamma_V$ is the width of the $V$ meson at $\mpp=M_V$,
\mbox{$p_\pi(M_{\pi\pi})=1/2\,\sqrt{M_{\pi\pi}^2-4M_\pi^2}$} is the
pion momentum in the $\pi^+\pi^-$ centre-of-mass frame, $p_\pi(M_V)$
is the pion momentum in the $V$-meson rest frame, and $M_\pi$ is the
pion mass.

\section{Experimental set-up}

The analyzed data were collected with the ZEUS detector at the HERA
collider in the years 1998--2000, when 920$\gevu$ protons collided with
27.5$\gevu$ electrons or positrons. The sample used for this study
corresponds to 81.7 pb$^{-1}$ of which 65.0 pb$^{-1}$ were collected
with an $e^+$ and the rest with an $e^-$ beam\footnote{From now on,
electrons and positrons will be both referred to as electrons in this
paper.}.

A detailed description of the ZEUS detector can be found
elsewhere~\cite{bluebook}. A brief outline of the components
that are most relevant for this analysis is given below.

The charged particles were tracked in the central tracking detector
(CTD)~\cite{CTD} which operated in a magnetic field of $1.43\Tesla$
provided by a thin superconducting solenoid. The CTD consisted of
72~cylindrical drift chamber layers, organized in nine superlayers
covering the polar-angle\footnote{The ZEUS coordinate system is a
right-handed Cartesian system, with the $Z$ axis pointing in the
proton beam direction, referred to as the ``forward direction'', and
the $X$ axis pointing left towards the centre of HERA.  The coordinate
origin is at the nominal interaction point.  The polar angle,
$\theta$, is measured with respect to the proton beam
direction.\xspace} region
\mbox{$15^\circ<\theta<164^\circ$}. The transverse-momentum
resolution for full-length tracks was $\sigma(p_T)/p_T = 0.0058p_T
\oplus 0.0065 \oplus 0.0014/p_T$, with $p_T$ in$\gevu$. 

The scattered electron was identified in the high-resolution
uranium--scintillator calorimeter (CAL)~\cite{CAL} which covered
99.7\% of the total solid angle and consisted of three parts: the
forward (FCAL), the barrel (BCAL) and the rear (RCAL)
calorimeters. Each part was subdivided transversely into towers and
longitudinally into one electromagnetic section (EMC) and either one
(in RCAL) or two (in BCAL and FCAL) hadronic sections (HAC).  The CAL
energy resolution, as measured under test-beam conditions, was
$\sigma(E)/E=0.18/\sqrt{E}$ for electrons and
$\sigma(E)/E=0.35/\sqrt{E}$ for hadrons, with $E$ in$\gevu$.

The position of the scattered electron was determined by combining
information from the CAL, the small-angle rear tracking
detector~\cite{srtd} and the hadron-electron
separator~\cite{hes}.

The luminosity was measured from the rate of the bremsstrahlung
process $ep\rightarrow e\gamma\:p$. The photon was measured in a
lead--scintillator calorimeter~\cite{lumi1,lumi2,lumi3} placed in the
HERA tunnel at $Z$ = $-$107 m.

\section{Data selection and reconstruction}

The online event selection required an electron candidate in the CAL,
along with the detection of at least one and not more than six tracks
in the CTD.

In the offline selection, the following further requirements were imposed:

\begin{itemize}
\item the presence of a scattered electron, with energy in the CAL
  greater that 10$\gevu$ and with an impact point on the face of the
  RCAL outside a rectangular area of \mbox{26.4 $\times$ 16 cm$^2$} in
  the $X-Y$ plane;
\item the $Z$ coordinate of the interaction vertex was within $\pm$ 50 cm of 
the nominal interaction point;
\item in addition to the scattered electron, the presence of exactly two 
oppositely charged tracks. Both tracks have to be associated with the
reconstructed vertex, each having pseudorapidity $|\eta|$ less than
1.75 and transverse momentum greater that 150 \mev. This ensures high
reconstruction efficiency and excellent momentum resolution in the
CTD.  These tracks were treated in the following analysis as a
$\pi^+\pi^-$ pair;
\item $E - P_Z > 45\gevu$, where $E- P_Z = \sum_i(E_i - P_{Z_i})$ and
  the summation is over the energy $E_i$ and longitudinal momentum
  $P_{Z_i}$ of the final-state electron and pions. This cut excludes
  events with high-energy photons radiated in the initial state;
\item events with any energy deposit larger than 300 MeV in the CAL,  not 
associated with the pion tracks (so-called \textquoteleft{unmatched}
islands'), were rejected.
\end{itemize}

The following kinematic variables are used to describe the exclusive 
production of a $\pi^+\pi^-$ pair:
\begin{itemize}
\item $Q^2$, the four-momentum squared of the virtual
photon; 
\item $W^2$, the squared centre-of-mass energy of the photon-proton
  system;
\item $\mpp$, the invariant mass of the two pions;
\item $t$, the squared four-momentum transfer at the proton vertex;
\item $\Phi_h$, the angle between the $\pi^+\pi^-$ production plane and the 
positron scattering plane in the $\gamma^\ast p$ centre-of-mass frame;
\item $\theta_h$ and $\phi_h$, the polar and azimuthal angles of the
  positively charged pion in the $s$-channel helicity frame~\cite{sw}
  of the $\pi^+\pi^-$.
\end{itemize}

The kinematic variables were reconstructed using the so-called
\textquoteleft{constrained}' method~\cite{constraint}, which uses 
the momenta of the decay particles measured in the CTD and the
reconstructed polar and azimuthal angles of the scattered electron.
The analysis was restricted to the kinematic region
$2<Q^2<80\gevu^2$, $32<W<180\gevu$, $|t|\le 0.6\gevu^2$ and 
$0.4< M_{\pi\pi} <2.5\gevu$. The lower mass range excludes
reflections from the $\phi \to K^+K^-$ decays and the upper limit
excludes the $J/\psi \to \mu^+\mu^-, e^+e^-$ decays with its
radiative tail.

The above selection yielded 63517 events for this analysis. 

The above cuts do not eliminate events in which the proton dissociates
into a low-mass final state, the products of which disappear down
the beam pipe. This contribution, estimated~\cite{zrho} to be about
20\% in the range of this analysis, was found to be $Q^2$ and $W$
independent. Its presence does not affect the conclusions of this
analysis.

\section{ Monte Carlo simulation} 

The program \textsc{Zeusvm}~\cite{muchor} interfaced to
\textsc{Heracles4.4}~\cite{heracl} was used. The effective
distributions of $Q^2, \ W$ and $|t|$ were parameterized to reproduce
the data. The mass and angular distributions were generated uniformly
and the MC events were then iteratively reweighted using the results
of the analysis.

The generated events were passed through a full simulation of the ZEUS
detector based on \textsc{Geant} 3.21~\cite{geant} and processed
through the same chain of selection and reconstruction procedures as
the data, accounting for trigger as well as detector acceptance and
smearing effects.  The number of simulated events after reconstruction
was approximately seven times greater than the number of reconstructed
data events.

\begin{figure}[tp]
\begin{center}
\epsfig{file=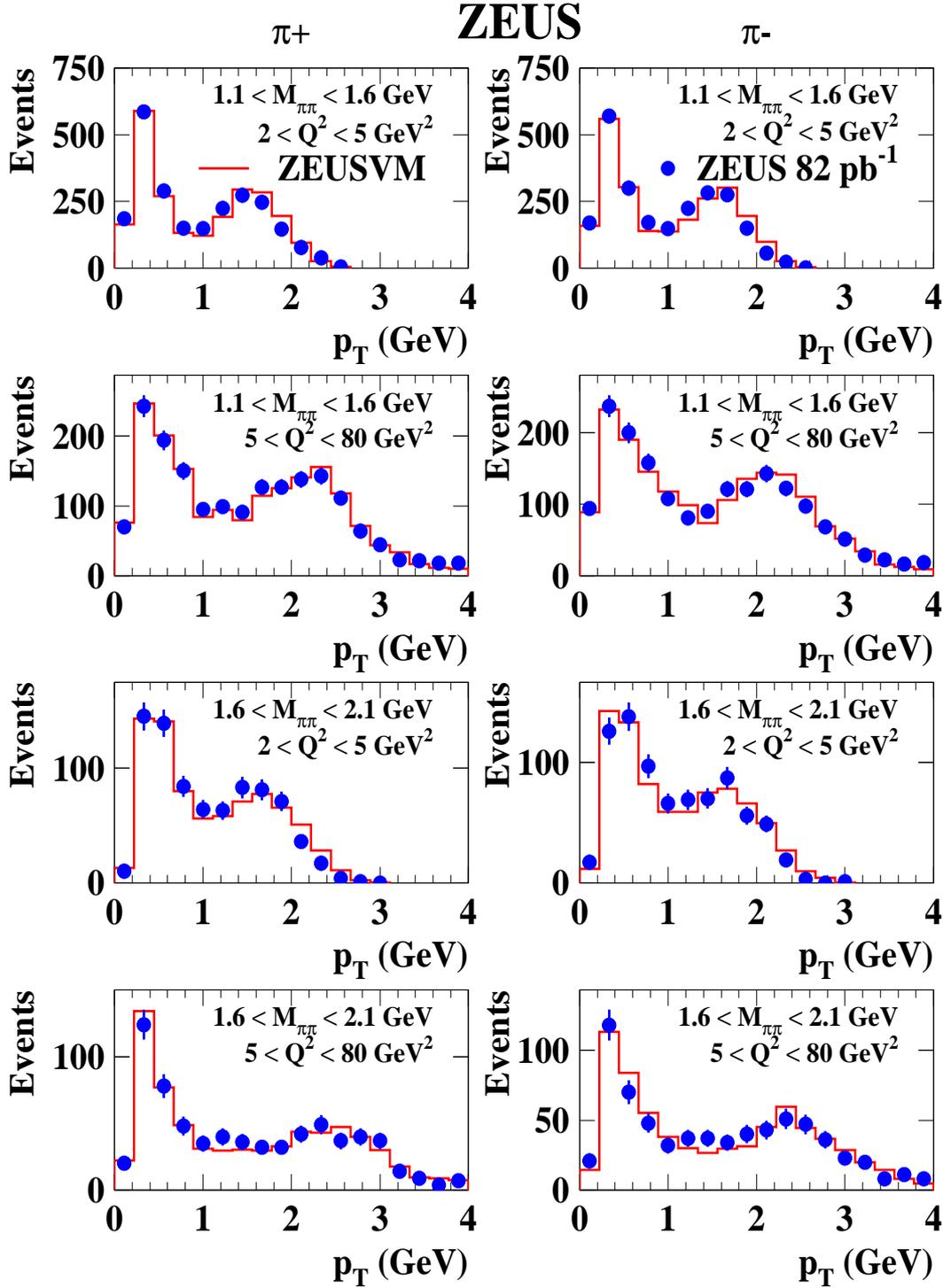,
     height=20.0cm,width=15.0cm}
\caption{Comparison between the data and the \textsc{Zeusvm} MC distributions 
for the transverse momentum, $p_T$, of $\pi^+$ and $\pi^-$ particles 
for different ranges of $Q^2$ and $M_{\pi\pi}$ as indicated in the figure.
The MC distributions are normalized to the data.}
\label{fig:comp1}
\end{center}
\end{figure}

\begin{figure}[tp]
\begin{center}
\epsfig{file=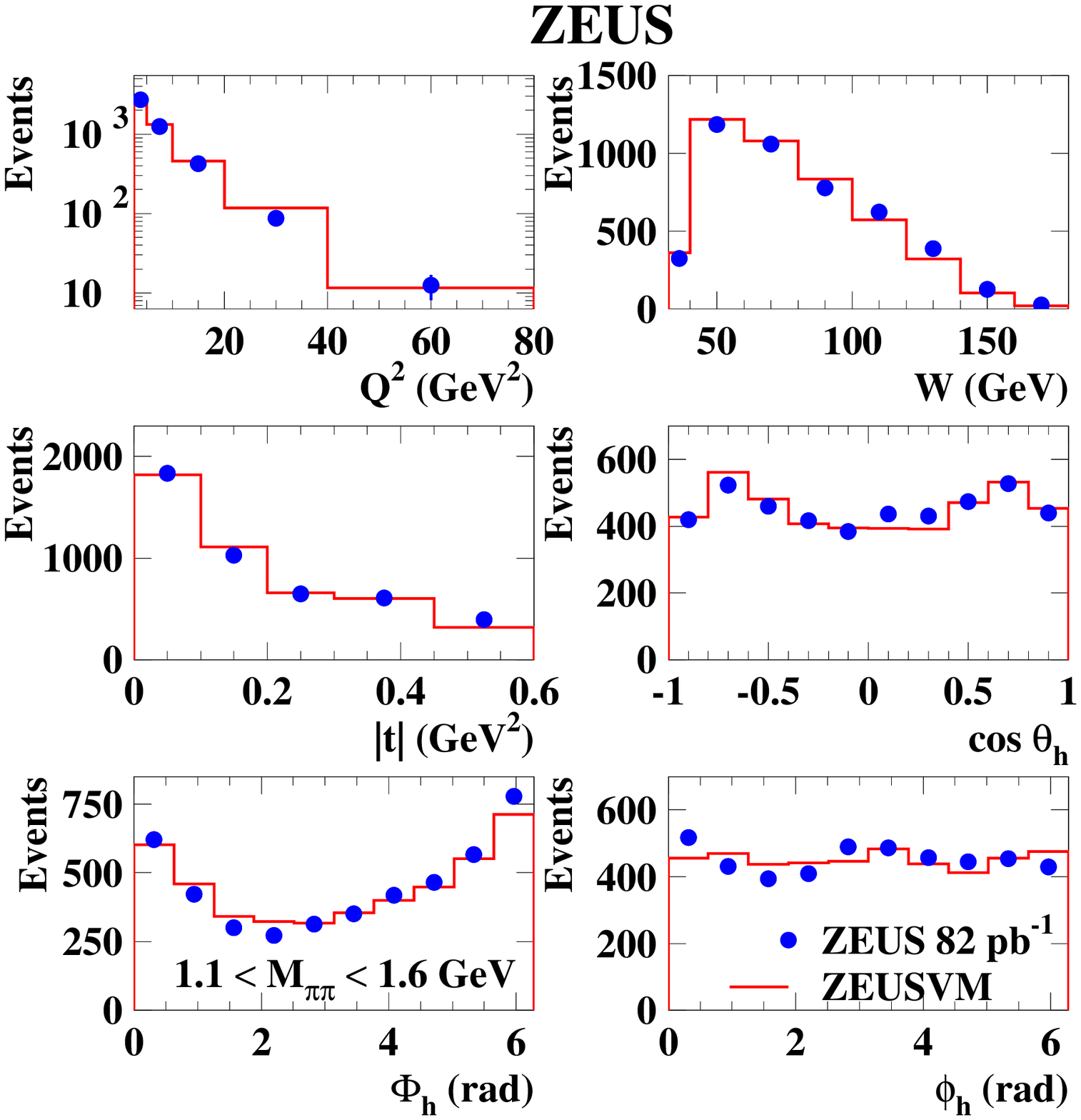,
     height=15.0cm,width=15.0cm}
\caption{Comparison between the data and the \textsc{Zeusvm} 
MC distributions for $Q^2, \ W$, $|t|$, $\cos\theta_h$, $\Phi_h$ and
$\phi_h$ for events within mass range $1.1<M_{\pi\pi}<1.6$ GeV. The MC
distributions are normalized to the data.}
\label{fig:comp3}
\end{center}
\end{figure}

\begin{figure}[tp]
\begin{center}
\epsfig{file=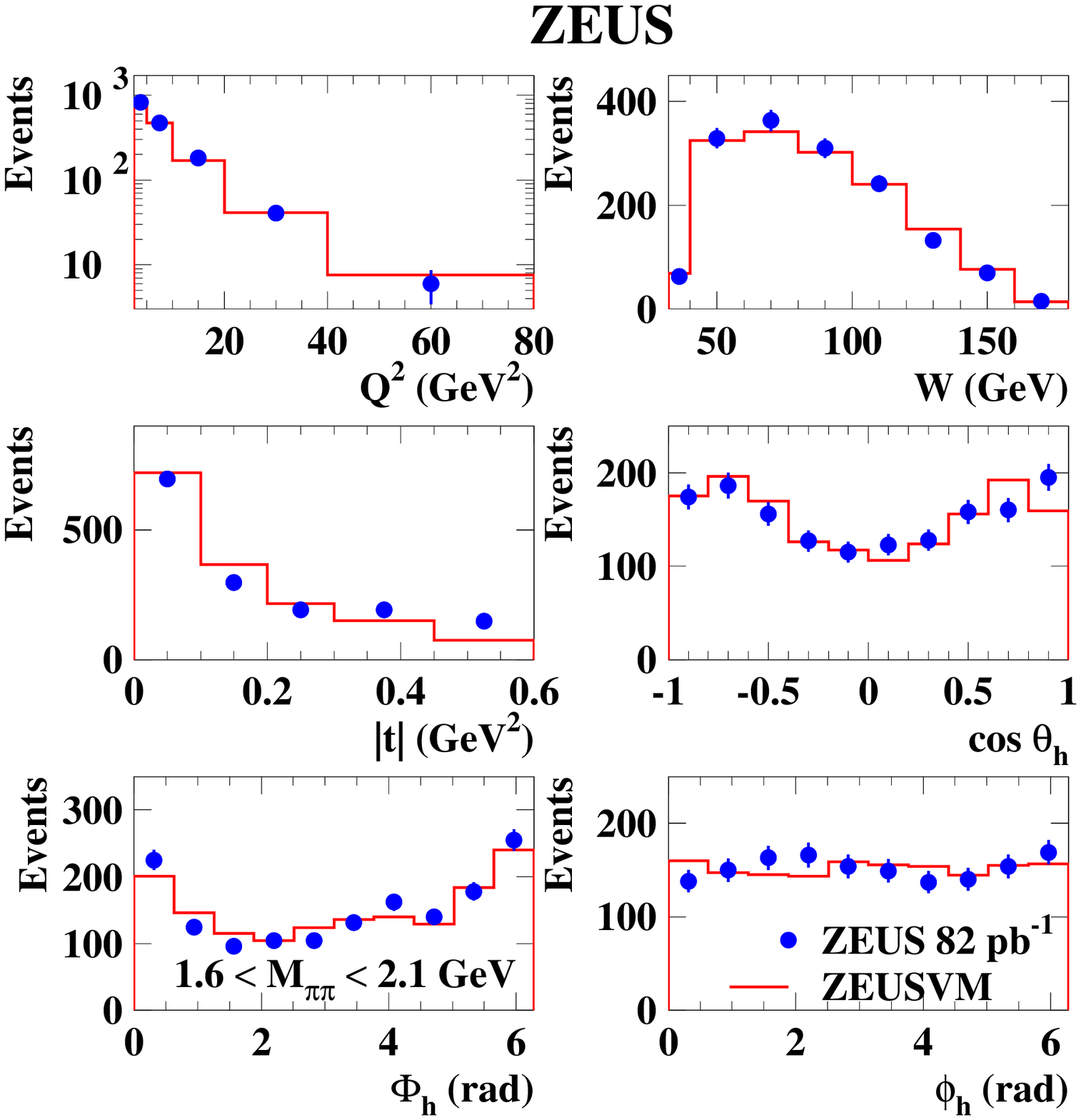,
     height=15.0cm,width=15.0cm}
\caption{Comparison between the data and the \textsc{Zeusvm} 
MC distributions for $Q^2, \ W$, $|t|$, $\cos\theta_h$, $\Phi_h$ and
$\phi_h$ for events within mass range $1.6<M_{\pi\pi}<2.1$ GeV. The MC
distributions are normalized to the data.}
\label{fig:comp5}
\end{center}
\end{figure}

A detailed comparison between the data and the \textsc{Zeusvm} MC
distributions for the mass range \mbox{$0.65<\mpp<1.1\gevu$}  has been
presented elsewhere~\cite{zrho}.  Some examples for the mass range
$1.1<M_{\pi\pi}<2.1\gevu$  are shown here. The transverse momentum,
$p_T$, of the $\pi^+$ and the $\pi^-$ particles for different ranges
of $Q^2$ and $M_{\pi\pi}$ are presented in
Fig.~\ref{fig:comp1}. Figure~\ref{fig:comp3} shows the $Q^2, \ W, \
|t|$, $ \cos \theta_h, \phi_h,$ and $\Phi_h$ distributions for events
selected within the mass ranges $1.1<M_{\pi\pi}<1.6$ $\gevu$, while
Fig.~\ref{fig:comp5} shows those distributions for the mass range
$1.6<M_{\pi\pi}<2.1\gevu$. All measured distributions are well
described by the MC simulations.

\section{The $\boldmath{\pi\pi}$ mass fit}

\begin{figure}
\begin{center}
\epsfig{file=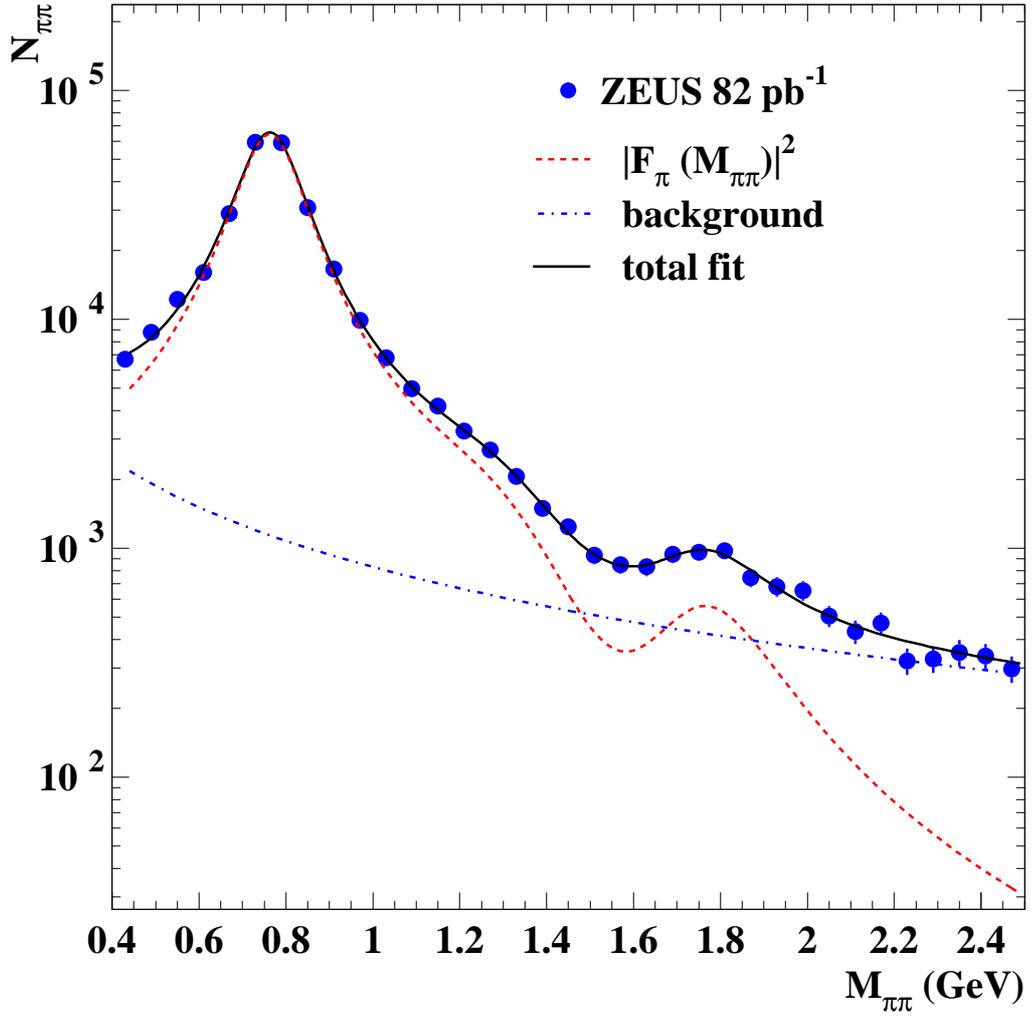,
     height=15.0cm,width=15.0cm}
\caption{The two-pion invariant-mass distribution, $M_{\pi\pi}$, 
where $N_{\pi\pi}$ is the acceptance-corrected number of events in
each bin of 60 MeV. The dots are the data and the full line is the
result of a fit using the Kuhn-Santamaria parameterization. The dashed
line is the result of the pion form factor normalized to the data and
the dash-dotted line denotes the background contribution.}
\label{fig:main-fit}
\end{center}
\end{figure}

The $\pi^+\pi^-$ mass distribution, after acceptance correction
determined from the above MC simulation, is shown in
Fig.~\ref{fig:main-fit}. A clear peak is seen in the $\rho$ mass
range. A small shoulder is apparent around 1.3$\gevu$ and a secondary
peak at about 1.8$\gevu$.

\begin{figure}
\begin{center}
\epsfig{file=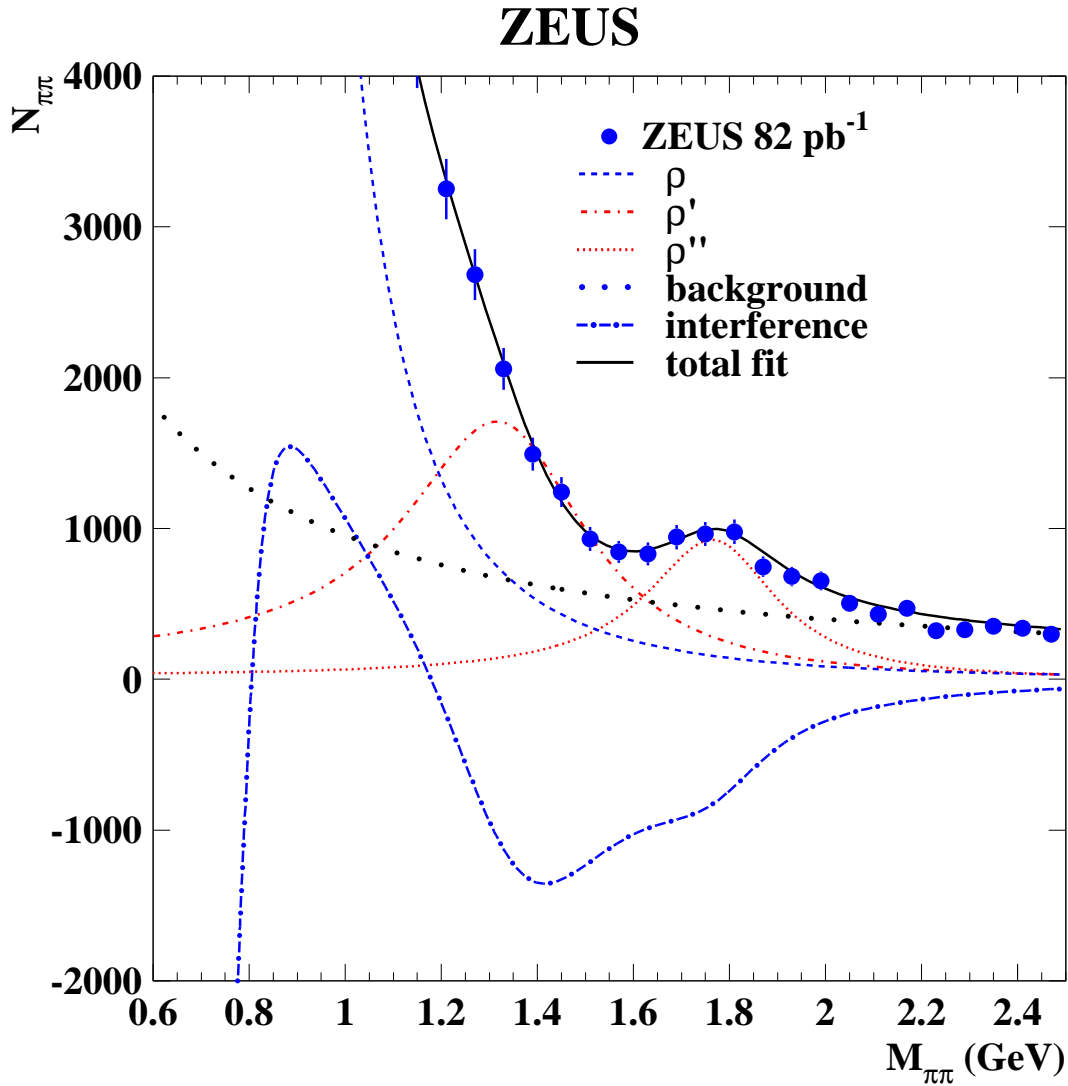,
     height=15.0cm,width=15.0cm}
\caption{The two-pion invariant-mass distribution, $M_{\pi\pi}$, where 
$N_{\pi\pi}$ is the acceptance corrected number of events in each bin
of 60 MeV. The dots are the data and the full line is the result of a
fit using the Kuhn-Santamaria parameterization. The contributions of the
three resonances $\rho$, $\rhp$ and $\rhpp$ are shown as dashed,
dash-dotted and dotted lines, respectively. The sum of their
interferences is shown by the long-dash-dotted line. The background is
presented as the sparse dotted line.}
\label{fig:lost-peak}
\end{center}
\end{figure}

The two-pion invariant-mass distribution was fitted, using the
least-square method~\cite{minuit}, as a sum of two terms,
\begin{equation}
\frac{dN(\mpp)}{d\mpp}=A\left(1-\frac{4M_\pi^2}{M_{\pi\pi}^2}\right)
\Biggl[\ffpi  +  {B}\left(\frac{M_0}{\mpp}\right)^n \Biggr],
\label{dndm}
\end{equation}
where $A$ is an overall normalization constant. The second term is a
parameterization of the non-resonant background, with constant
parameters $B$, $n$ and $M_0$ = 1$\gevu$. The other parameters, the
masses and widths of the three resonances and their relative
contributions $\beta$ and $\gamma$, enter through the pion form
factor, $F_\pi$ (Eq.~\ref{fpion}). The fit, which includes 11
parameters, gives a good description of the data
($\chi^2$/ndf=28.8/24=1.2). The result of the fit is shown in
Fig.~\ref{fig:main-fit} together with the contribution of each of the
two terms of Eq.~\ref{dndm}. The $\rho$ and the $\rhpp$ signals are
clearly visible. The negative interference between all the resonances
results in the $\rhp$ signal appearing as a shoulder.  To illustrate
this better, the same data and fit are shown in
Fig.~\ref{fig:lost-peak} on a linear scale and limited to $M_{\pi\pi} >$
1.2$\gevu$, with separate contributions from the background, the three
resonant amplitudes as well as their total interference term.

The fit parameters are listed in Table~\ref{table1}. Also listed are
the mass and width parameters from the Particle Data Group (PDG)
~\cite{pdg}.  The masses and widths of the $\rh$ and the $\rhpp$ as
well as the width of the $\rhp$ agree with those listed in the PDG,
while there is about 100 MeV difference between the PDG value and the
fitted mass of the $\rhp$. It should however be noted that the value
quoted by PDG is an average over many measurements having a large
spread ($1265 \pm 75$ up to $1424 \pm 25$ MeV for the $\pi\pi$ decay
mode) in this mass range.
 
The measured negative value of $\beta$ and positive value of $\gamma$
implies that the relative signs of the amplitudes of the three
resonances $\rho,
\rhp$ and $\rhpp$ are $+, -, +$, respectively.  A similar pattern
was observed in $e^+e^- \to \pi^+\pi^-$ and $\tau$-decay experiments
~\cite{dm2,cmd2,cmd2-a,cmd2-b,snd,babar,aleph,cleo,belle10}, which also
showed a dip in the mass range around 1.6$\gevu$, resulting from 
destructive interference. There is a single experiment where a
constructive interference was obtained around 1.6 $\gevu$, namely $\gamma p
\to \pi^+ \pi^- p$~\cite{aston}, a result which is not
understood~\cite{kdp99}.

In the mass fits above it was assumed that the relative amplitudes
$\beta$ and $\gamma$ are real.  In order to test this assumption, 
the fit was repeated allowing them to be complex.
The pion form factor was re-written in the form
\begin{equation}
\fpi=\frac{BW_{\rh}(\mpp) + \beta_0\cdot \exp(i\Phi_{12}) BW_{\rhp}(\mpp) + 
\gamma_0\cdot 
\exp(i\Phi_{13}) BW_{\rhpp}(\mpp)}{1 + \beta_0 + \gamma_0},
\label{complex}
\end{equation}
\noindent
where $\beta_0$ and $\gamma_0$ are real numbers and two
additional fit parameters, $\Phi_{12}$ and $\Phi_{13}$, are the
corresponding phase shifts. The value of the phase-shifts obtained
from the fit were $\Phi_{12}=3.2\pm0.2$ rad and $\Phi_{13}=0.1\pm0.2$
rad, supporting the assumption of the real nature of the relative
amplitudes.

\section{Systematic uncertainties}

The systematic uncertainties of the fit parameters were evaluated by
varying the selection cuts and the MC simulation
parameters. Motivation for the variation in cuts used below can be
found in a previous ZEUS analysis~\cite{zrho}. The following selection
cuts were varied:
\begin{itemize}
\item the $E - P_Z$ cut was changed within the resolution 
of $\pm$ 3$\gevu$;
\item the $p_T$ threshold for the pion tracks (default 0.15$\gevu$) was 
increased to 0.2$\gevu$ and the $|\eta|$ cut on the two pion tracks was
changed (default 1.75) by $\pm 0.25$;
\item the required maximum distance of closest approach of the two 
extrapolated pion tracks to the matched island in the CAL was changed
from 30 cm to 20 cm;
\item the $Z$-vertex cut was varied by $\pm 10$ cm;
\item  the energy threshold for an unmatched island (elasticity cut) was 
changed by $\pm$50~\mev;
\item the bin size in the fitted mass distribution  (default 60 MeV) was 
varied by \mbox{$\pm$ 20 MeV};
\item the mass range was narrowed to  $0.5<M_{\pi\pi}<2.3\gevu$;
\item the $|t|$ cut  was varied by $\pm 0.1\gevu^2$; 
\item the $W$ range was changed to $35<W<190\gevu$;
\item the $\cos \theta_h$ range was changed to $|\cos \theta_h| < 0.9$;
\item the $W^\delta$ dependence in the MC was varied by changing the 
$Q^2$-dependent $\delta$ value by $\pm$ 0.03;
\item the exponential $t$ distribution in the MC was reweighted by changing 
the nominal $Q^2$-dependent slope parameter $b$ by $\pm$ 0.5$\gevu^{-2}$;
\item the exponent of the $Q^2$ distribution parameterization in the MC was 
changed by $\pm$0.05.
\end{itemize}

The largest variations were observed for $\gamma$, $\Gamma(\rhpp)$ and
$\beta$. The value of $\Gamma(\rhpp)$ changes by 7\% when the
elasticity cut is varied. The restriction of the phase space in the
fitted mass range leads to a change of the value of $\beta$ by
$-5.2\%$ while for $\gamma$, restricting the $|\cos \theta_h|$ range
leads to a change of $-8\%$. In addition, another form of background
in Eq.~\ref{dndm}, with an added exponential term, was
investigated. It gave a very similar result in the mass range of this
analysis and therefore no additional uncertainty was assigned to the
form of the fitted mass curve.

All the systematic uncertainties were added in quadrature. The
combined systematic uncertainties are included in Table~\ref{table1}.

\section{Decay angular distributions}

Decay angular distributions can be used to determine the spin
density-matrix elements of a resonance~\cite{sw,zspin}. In the present
case we study three resonances, all in a $J^P = 1^-$ state. However,
the decay angular distribution in a given mass bin is affected by the
background contribution which does not necessarily have the same
quantum numbers as the resonance. Given the above, only the
distribution of the polar angle $\theta_h$, defined as the polar angle
of the positively charged pion in the helicity frame, was studied.

\begin{figure}
\begin{center}
\epsfig{file=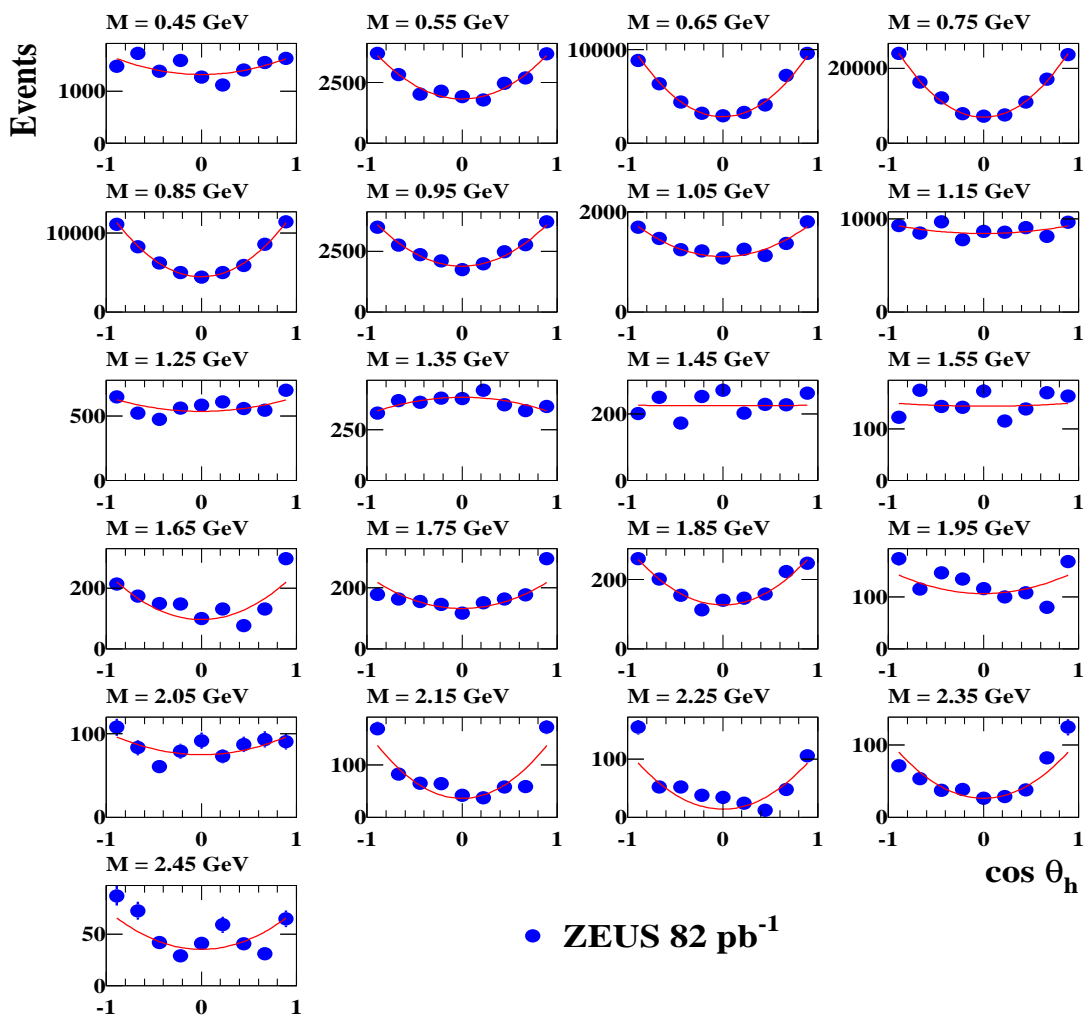,
     height=15.0cm,width=15.0cm}
\caption{ The acceptance-corrected $\cos\theta_h$ distribution for 
different $M_{\pi\pi}$ intervals, with the mean mass values indicated in the
figure. The lines represent fits to the data as discussed in the text.}
\label{fig:hel-bin1} 
\end{center}
\end{figure}

\begin{figure}
\begin{center}
\epsfig{file=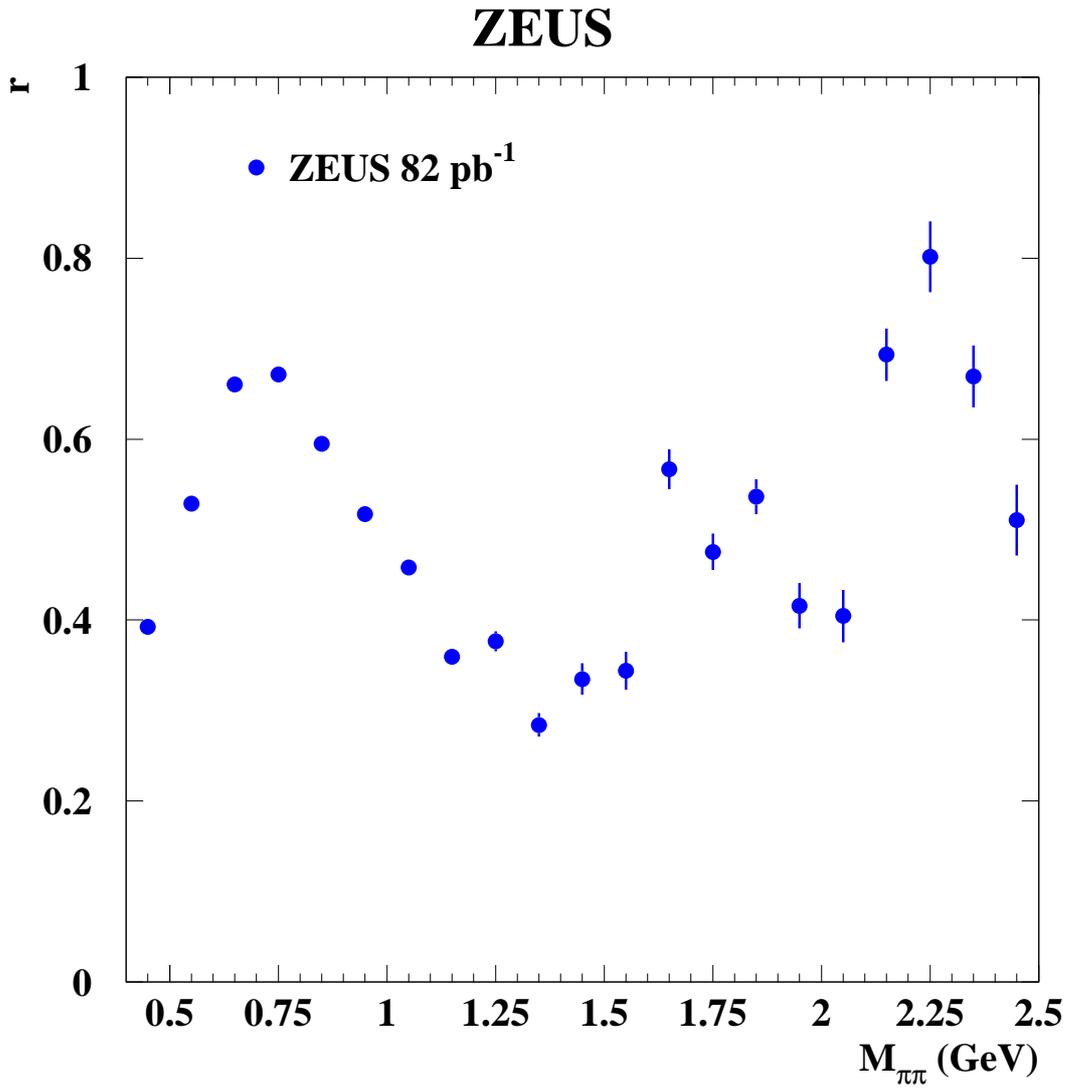,
     height=15.0cm,width=15.0cm}
\caption{ The fitted parameter $r$ as a function of the two-pion invariant 
mass, $M_{\pi\pi}$. Only statistical uncertainties are shown.}
\label{fig:long-tot}
\end{center}
\end{figure}

The distribution of $\cos\theta_h$ is shown in Fig.~\ref{fig:hel-bin1}
for different mass bins; its shape is clearly mass dependent. In order
to study the mass dependence further, the angular distribution of the
polar helicity angle, $W(\cos\theta_h)$ was parameterized as
\begin{equation}
W(\cos\theta_h)\propto [1 - r + (3r - 1)\cos^2\theta_h],
\label{angl}
\end{equation}
and fitted to the data.  The mass dependence of the resulting
parameter $r$ is shown in Fig.~\ref{fig:long-tot}. In the mass range
$M_{\pi\pi}<$ 1.1$\gevu$, $r$ shows the dependence seen for the
$r_{00}^{04}$ density matrix in the $\rho$ region~\cite{zrho}. Indeed
this region is dominated by exclusive production of $\rho$ and
therefore $r = r_{00}^{04}$. In that case, $r$ can be interpreted as
$\sigma_L/\sigma_{\rm tot}$, assuming $s$-channel helicity conservation
(SCHC). Here $\sigma_L$ is the cross section for producing $\rho$ by a
longitudinally polarized photon, and $\sigma_{tot} = \sigma_L +
\sigma_T$, with $\sigma_T$ the production cross section by
transversely polarized photons. The results shown here for the $\rho$
region are in excellent agreement with the values given
in an earlier ZEUS paper~\cite{zrho}.

The structure seen for $M_{\pi\pi}>$ 1.1$\gevu$  is not easy to
interpret, however the dip observed around 1.3$\gevu$  and the
enhancement at 1.6$\gevu$  seem to follow the location of the resonances
determined from the mass distribution.

\section{$Q^2$ dependence of the pion form factor}

\begin{figure}
\begin{center}

\epsfig{file=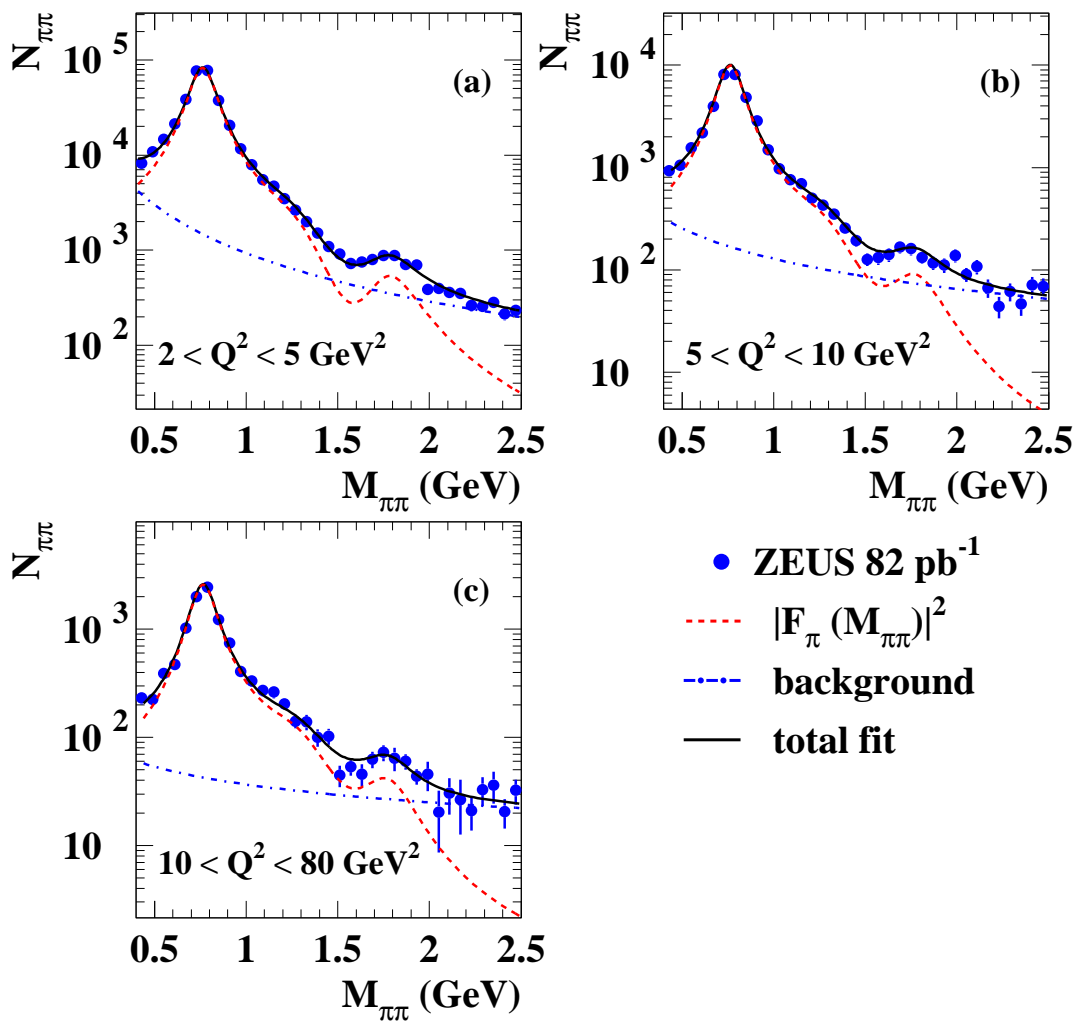,
     height=15.0cm,width=15.0cm}
\caption{The two-pion invariant-mass distribution, $M_{\pi\pi}$, 
where $N_{\pi\pi}$ is the acceptance-corrected number of events in
each bin of 60 MeV, for three regions of $Q^2$, as denoted in the
figure. The dots are the data and the full line is the result of a
fit using the Kuhn-Santamaria parameterization. The dashed line is the
result of the pion form factor normalized to the data and the
dash-dotted line denotes the background contribution.}
\label{fig:q2bin-3}
\end{center}
\end{figure}

The $Q^2$ dependence of the relative amplitudes was determined by
performing the fit to $M_{\pi\pi}$ in three $Q^2$ regions, 2--5, 5--10
and 10--80$\gevu^2$.  The masses and widths of the three resonances
were fixed to the values found in the overall fit and listed in
Table~\ref{table1}. The results are shown in Fig.~\ref{fig:q2bin-3}.
A reasonable description of the data is achieved in all three $Q^2$
regions. The corresponding values of $\beta$ and $\gamma$ are given in
Table~\ref{table2}. The absolute value of $\beta$ increases with $Q^2$
while the value of $\gamma$ is consistent with no $Q^2$ dependence,
within large uncertainties.

\begin{figure}
\begin{center}
\epsfig{file=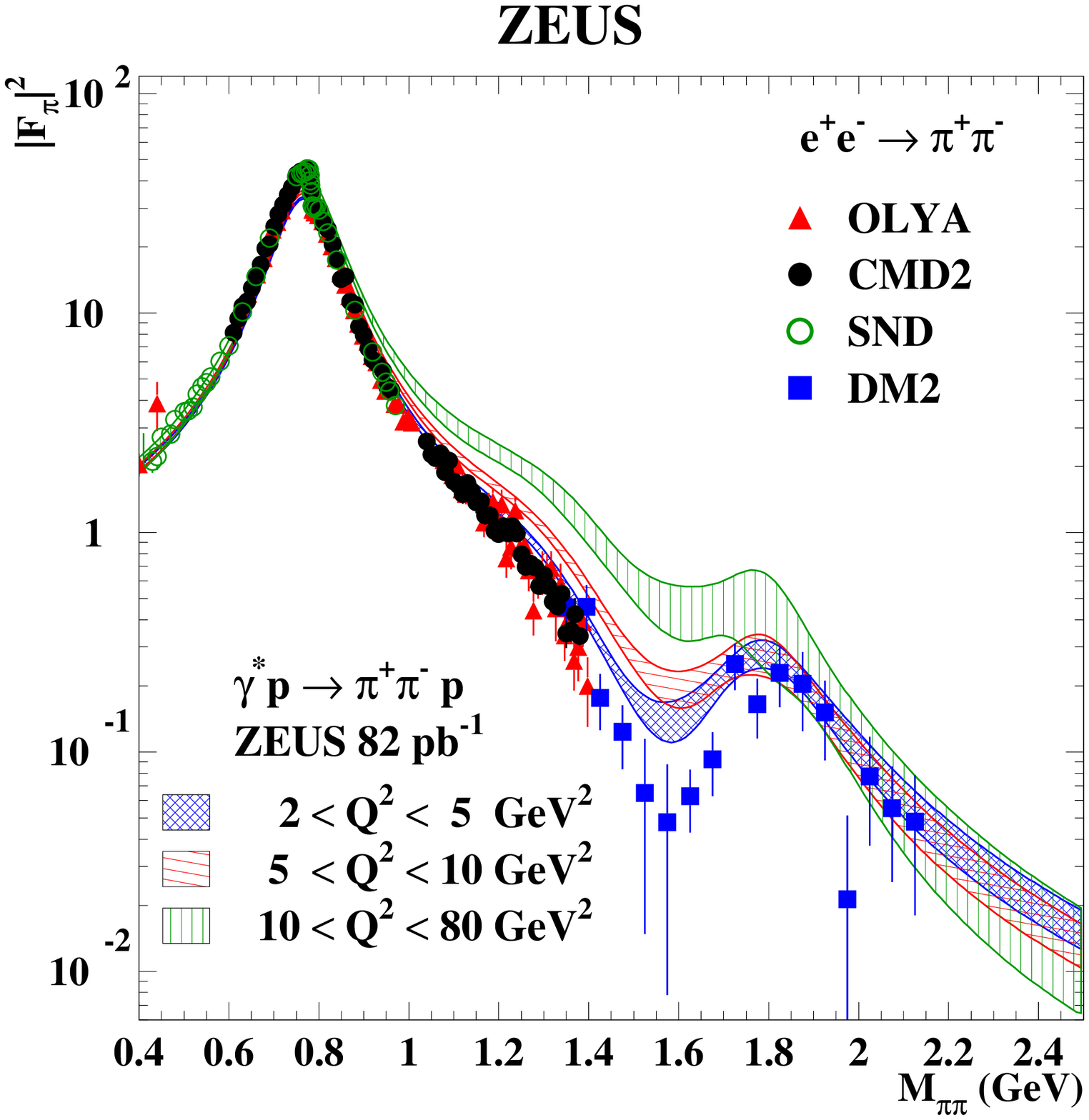,
     height=15.0cm,width=15.0cm}
\caption{The pion form factor squared, $|F_{\pi}|^2$, as a function of 
the $\pi^+\pi^-$ invariant mass, $M_{\pi\pi}$, as obtained from the
reaction
$e^+e^-\to\pi^+\pi^-$~\protect\cite{barkov,dm2,cmd2,cmd2-b,snd}. The
shaded bands represent the square of the pion form factor and its total
uncertainty obtained in the present analysis for three ranges of
$Q^2$: 2--5 GeV$^2$ (crossed lines), 5--10 GeV$^2$ (horizontal lines)
and 10--80 GeV$^2$ (vertical lines). }
\label{fig:comp-ee}
\end{center}
\end{figure}

\begin{figure}
\begin{center}
\epsfig{file=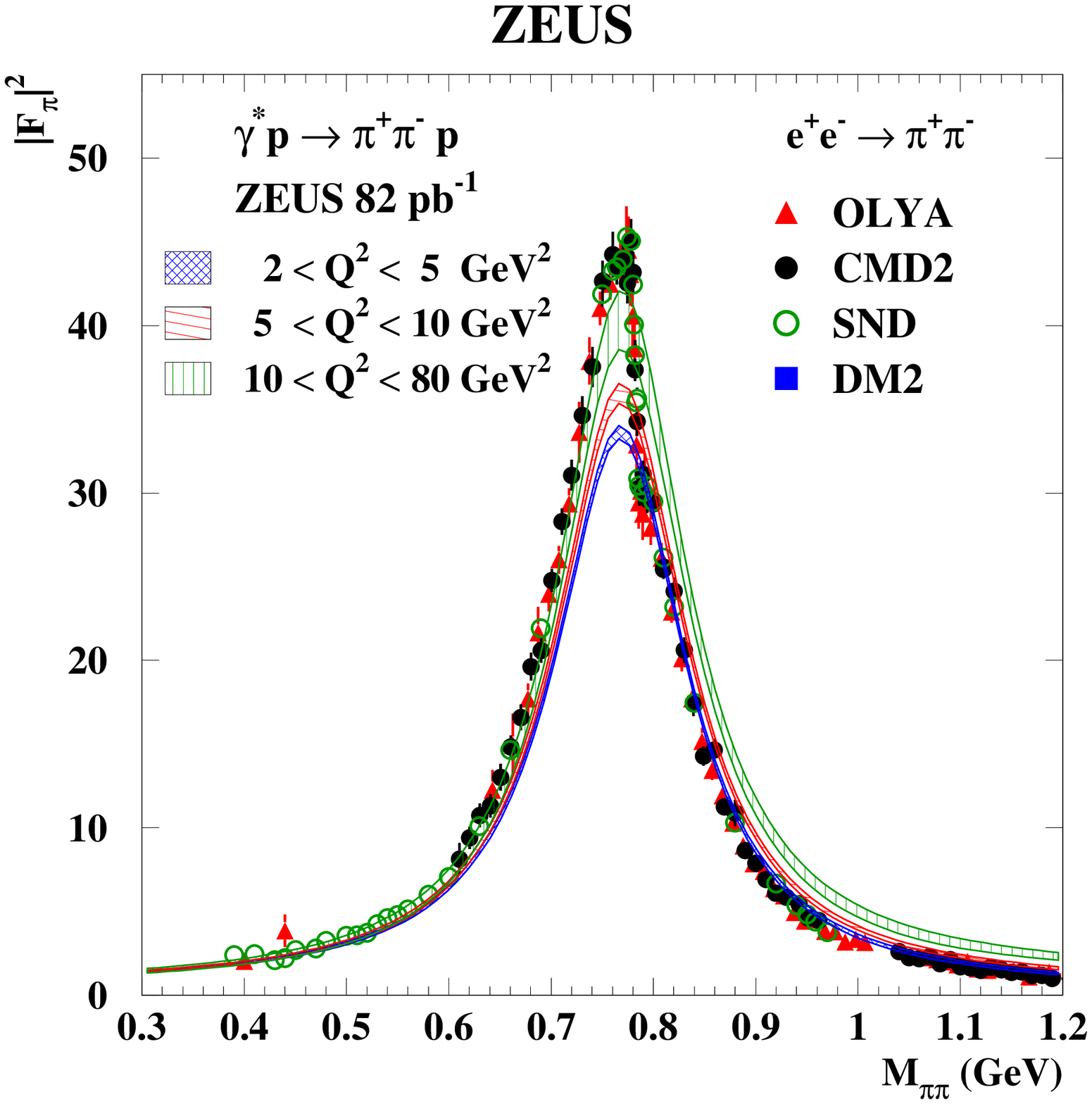,
     height=15.0cm,width=15.0cm}
\caption{
The pion form factor squared, $|F_{\pi}|^2$, in the $\rh$ mass region,
as a function of the $\pi^+\pi^-$ invariant mass, $M_{\pi\pi}$, as
obtained from the reaction
$e^+e^-\to\pi^+\pi^-$~\protect\cite{barkov,dm2,cmd2,cmd2-b,snd}. The
shaded bands represent the square of the pion form factor and its total
uncertainty obtained in the present analysis for three ranges of
$Q^2$: 2--5 GeV$^2$ (crossed lines), 5--10 GeV$^2$ (horizontal lines)
and 10--80 GeV$^2$ (vertical lines).  }
\label{fig:comp-ee-rho}
\end{center}
\end{figure}

Figure~\ref{fig:comp-ee} shows the curves representing the pion form
factor, $|F_{\pi}(M_{\pi\pi})|^2$, as obtained in the present analysis
for the three $Q^2$ ranges: 2--5, 5--10, 10--80$\gevu^2$. Also shown
are results obtained in the time-like regime from the reaction
$e^+e^-\to\pi^+\pi^-$.  In general, the features of the
$|F_\pi(M_{\pi\pi})|^2$ distribution observed here are also observed
in $e^+e^-$, i.e., the prominent $\rho$ peak, a shoulder around the
$\rhp$ and a dip followed by an enhancement in the $\rhpp$
region. Above the $\rho$ region, where the interference between the
$\rhp$ and the $\rhpp$ starts to dominate, there is a dependence of
$|F_\pi(M_{\pi\pi})|^2$ on $Q^2$, with the results from the lowest
$Q^2$ range closest to those from $e^+e^-$. However, in the region of
the $\rho$ peak, shown in Fig.~\ref{fig:comp-ee-rho}, the pion
form-factor $|F_\pi(M_{\pi\pi})|^2$ is highest at the highest $Q^2$,
as in the $\rhp$-$\rhpp$ interference region, while the $e^+e^-$ data
are higher than those in the highest $Q^2$ range. They are equal
within errors for $M_{\pi\pi} >$ 1.8$\gevu$.

\section{Cross-section ratios as a function of $Q^2$}

The $Q^2$ dependence of the $\rh$ by itself is given
elsewhere~\cite{zrho}. Since the $\pi\pi$ branching ratios of $\rhp$
and $\rhpp$ are poorly known, the ratio $R_V$ defined as
\begin{equation}  
R_V = \frac{\sigma(V)\cdot Br(V \to \pi\pi)}{\sigma(\rho)},
\end{equation}
has been measured, where $\sigma$ is the cross section for
vector-meson production and $Br(V \to \pi\pi)$ is the branching ratio
of the vector meson $V (\rhp,\rhpp)$ into $\pi\pi$.  The ratio $R_V$
may be directly determined from the results of the $M_{\pi\pi}$ mass
fit,
\begin{equation}
R_{\rhp} = \beta^2 \frac{I_{\rhp}}{I_\rho} \ \ \ \ R_{\rhpp} = \gamma^2
\frac{I_{\rhpp}}{I_\rho},
\end{equation}
where
\begin{equation}
I_V = \int_{2M_\pi}^{M_V+5\Gamma_V} {dM_{\pi\pi} |BW_V(\mpp)|^2},
\end{equation}
and the integration is carried out over the range $2M_\pi <
M_{\pi\pi} < M_V + 5 \Gamma_V$.

\begin{figure}
\begin{center}
\epsfig{file=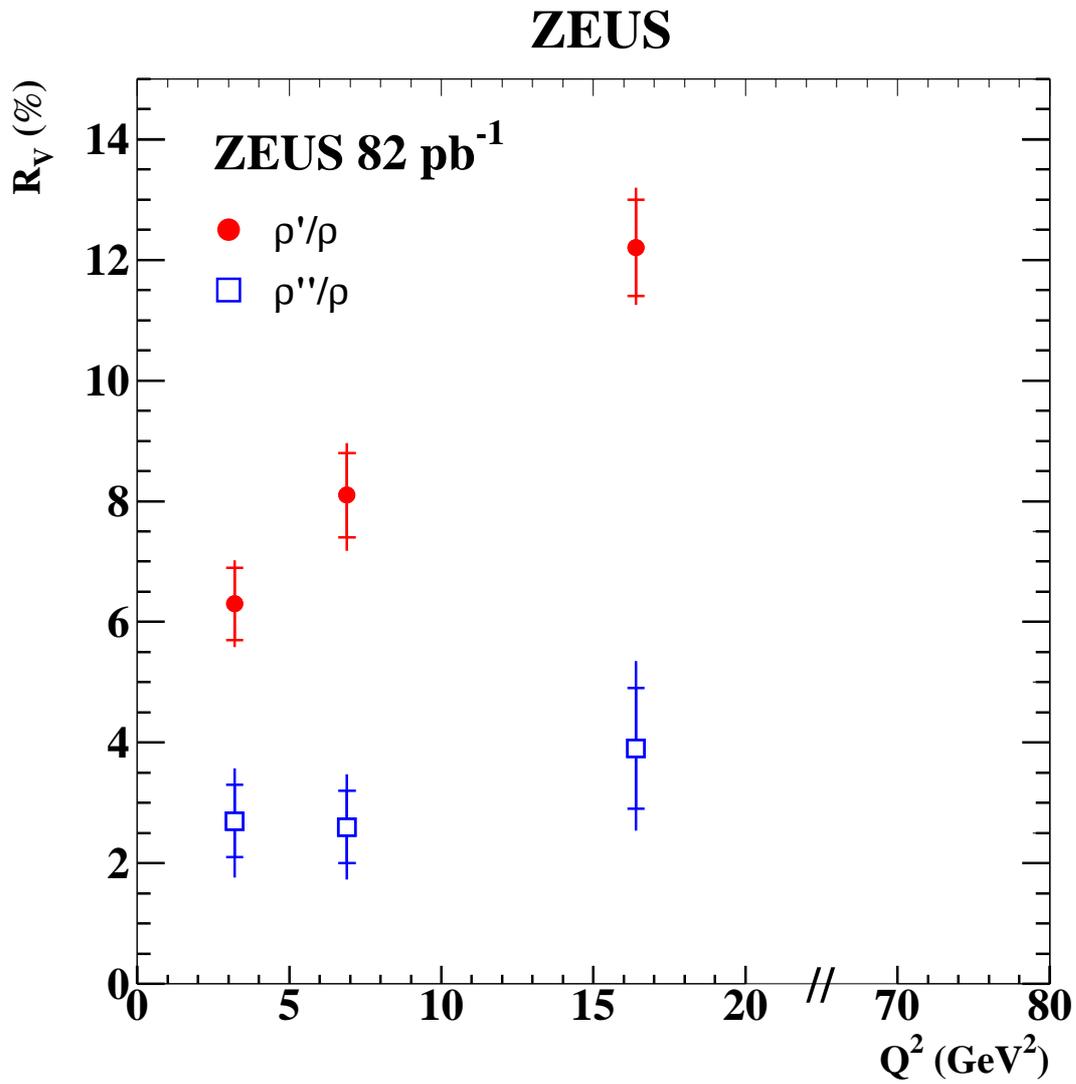,
     height=15.0cm,width=15.0cm}
\caption{The ratio $R_V$ as a function of $Q^2$ for $V = \rhp$ 
(full circles) and $\rhpp$ (open squares). The inner error bars
indicate the statistical uncertainty, the outer error bars represent
the statistical and systematic uncertainty added in quadrature.}
\label{fig:ratio}
\end{center}
\end{figure}

Figure~\ref{fig:ratio} shows and Table~\ref{table3} lists the ratio
$R_V$ for $V = \rhp, \rhpp$, as a function of $Q^2$.  Owing to the
large uncertainties of $R_{\rhpp}$, no conclusion on its $Q^2$
behaviour can be deduced, whereas $R_{\rhp}$ clearly increases with
$Q^2$.  This rise has been predicted by several
models~\cite{nnz94,fks96,in02,afs,iphd}.  The suppression of the $2S$
state ($\rhp$) is connected to a node effect which results in
cancellations of contributions from different impact-parameter regions
at lower $Q^2$, while at higher $Q^2$ the effect vanishes.

\section{ Summary}

Exclusive two-pion electroproduction has been studied by ZEUS at HERA
in the range $0.4<M_{\pi\pi}<2.5\gevu$, $2<Q^2<80\gevu^2$,
\mbox{$32<W<180\gevu$} and $|t| \le 0.6\gevu^2$.
The mass distribution is well described by the pion electromagnetic
form factor, $|F_\pi(M_{\pi\pi})|^2$, which includes three resonances,
$\rho$, $\rhp$(1450) and $\rhpp$(1700).

A $Q^2$ dependence of $|F_\pi(M_{\pi\pi})|^2$ is observed, visible in
particular in the interference region between $\rhp$ and $\rhpp$.  The
electromagnetic pion form factor obtained from the present analysis is
lower (higher) than that obtained from $e^+e^-\to\pi^+\pi^-$ for
$M_{\pi\pi} < 0.8\gevu$ ($0.8 < M_{\pi\pi} < 1.8\gevu$). They are
equal within errors for $M_{\pi\pi} > 1.8\gevu$.

The $Q^2$ dependence of the cross-section ratios $R_{\rhp} =
\sigma(\rhp\to\pi\pi)/\sigma(\rh)$ and $R_{\rhpp} =
\sigma(\rhpp\to\pi\pi)/\sigma(\rh)$, has been studied.  
The ratio $R_{\rhp}$ rises strongly with $Q^2$, as expected in
QCD-inspired models in which the wave-function of the vector meson is
calculated within the constituent quark model, which allows for nodes
in the wave-function to be present.

\section*{Acknowledgments}

It is a pleasure to thank the DESY Directorate for their strong
support and encouragement. The remarkable achievements of the HERA
machine group were essential for the successful completion of this
work and are greatly appreciated. The design, construction and
installation of the ZEUS detector has been made possible by the
efforts of many people who are not listed as authors.


\newpage
\clearpage

\begin{table}
  \begin{center}
    \begin{tabular}{|l|c|c|} \hline
  Parameter &   ZEUS    &  PDG \\ \hline
  M$_\rho$ (MeV) & $771\pm2^{+2}_{-1}$     & 775.49$\pm$0.34 \\ \hline
  $\Gamma_\rho$ (MeV) & $155\pm5\pm2$                & 149.1$\pm$0.8 \\ \hline
  $\beta$        & $ -0.27\pm0.02\pm0.02$       &             \\ \hline
  M$_{\rho^\prime}$ (MeV)& $1350\pm20^{+20}_{-30}$      & 1465$\pm$25 \\ \hline
  $\Gamma_{\rho^\prime}$ (MeV) &$ 460\pm30^{+40}_{-45}$ & 400$\pm$60 \\ \hline
  $\gamma$             & $0.10\pm0.02^{+0.02}_{-0.01}$    &  \\   \hline
  M$_{\rho^{\prime\prime}}$ (MeV)&$1780\pm20^{+15}_{-20}$& 1720$\pm$20 
\\ \hline
  $\Gamma_{\rho^{\prime\prime}}$ (MeV)& $310\pm30^{+25}_{-35}$& 
250$\pm$100 \\ \hline
  $B$ & $0.41\pm0.03\pm0.07$&   \\ \hline
  $n$ & $1.30\pm0.06^{+0.18}_{-0.13}$&   \\ \hline
   \end{tabular}
\end{center}
\caption
{ Fit parameters obtained using $\fpi$ parameterization. Masses and
widths are in MeV. The first uncertainty is statistical, the second
systematic. Also shown are the masses and widths from the
PDG~\protect\cite{pdg}.}
\label{table1}
\end{table}

\newpage

\begin{table}
  \begin{center}
    \begin{tabular}{|l|c|c|c|} \hline
  $ Q^2 (\rm GeV^2)$   & 2--5   & 5--10  &  10--80 \\ \hline
  $ \beta$  & $-0.249\pm0.008^{+0.005}_{-0.003}$  & 
$-0.282\pm0.008^{+0.005}_{-0.008}$ & $-0.35\pm0.02\pm0.01$ \\ \hline
  $\gamma$ & $0.100\pm0.009\pm0.003$    & $0.098\pm0.012^{+0.005}_{-0.003}$ & 
$0.118\pm0.022^{+0.008}_{-0.006}$ \\ \hline
   \end{tabular}
\end{center}
\caption
 {The $Q^2$ dependence of the $\beta$ and $\gamma$  parameters.
Masses and widths are fixed to the values given in Table~\ref{table1}.
The first uncertainty is statistical, the second systematic.}
\label{table2}
\end{table}

\begin{table}
  \begin{center}
    \begin{tabular}{|l|c|c|c|} \hline
  $ Q^2 (\rm GeV^2)$   & 2--5   & 5--10  &  10--80 \\ \hline
  $ R_{\rhp}$  & 0.063$\pm0.006\pm0.004$  & 
0.081$\pm0.007^{+0.006}_{-0.005}$ & 0.122$\pm0.008^{+0.005}_{-0.006}$ \\ 
\hline
  $R_{\rhpp}$ & 0.027$\pm0.006^{+0.004}_{-0.003}$    
& 0.026$\pm0.006\pm0.003$ & 
0.039$\pm0.010^{+0.003}_{-0.005}$ \\ \hline
   \end{tabular}
\end{center}
\caption
 {The $Q^2$ dependence of the ratio $R_V$ for $V=\rhp$ and $\rhpp$.
The first uncertainty is statistical, the second systematic.}
\label{table3}
\end{table}


\newpage
\clearpage

\setlength{\unitlength}{1mm}

\newpage

\newpage

\newpage

\newpage

\newpage

\newpage

\newpage

\newpage

\newpage

\newpage

\end{document}